\documentclass[a4paper,11pt]{article}

\usepackage{amsmath}
\usepackage{amssymb}
\usepackage{amsthm}
\usepackage{array}
\usepackage{indentfirst}
\usepackage{graphicx}
\usepackage[utf8]{inputenc}
\usepackage[dvipsnames]{xcolor}
\usepackage{tabularx}
\usepackage{caption}
\usepackage[normalem]{ulem}
\usepackage{subcaption}
\usepackage{array}
\usepackage{amssymb}

\newcolumntype{Y}{>{\centering\arraybackslash}X}
\newcolumntype{W}[1]{>{\centering\arraybackslash\hsize=#1\hsize}X}

\usepackage{jcappub}
\definecolor{myblue}{rgb}{0.05,0.1,0.5}

\definecolor{alcolor}{rgb}{0.1,0.5,0.05}

\hypersetup{colorlinks=true, linkcolor=blue, urlcolor=blue, citecolor=blue, linktocpage=true}

\begin{document}
%%%%%%%%%%%%%%%%%%%%

\title{From raw data to neutrino candidates: a neural-network pipeline for Baikal-GVD}

\author[a, b]{A. Matseiko,}
\author[a]{G. Plotnikov,}
\author[a, b]{I. Kharuk.}

\affiliation[a]{Institute for Nuclear Research of the Russian Academy of Sciences,
\\ 60th October Anniversary Prospect 7a, Moscow, 117312, Russia}
\affiliation[b]{Moscow Institute of Physics and Technology,\\
Institutsky lane, 9, Dolgoprudny, Moscow region, 141700, Russia}

\emailAdd{ivan.kharuk@phystech.edu}
\emailAdd{matseiko.av@phystech.edu}
\emailAdd{plotnikov.gp@phystech.edu}

\abstract{
We present a neural-network-based data processing pipeline for Baikal-GVD, designed to improve event reconstruction quality and accelerate neutrino candidates selection. 
The pipeline comprises three stages: fast suppression of extensive air shower events, suppression of noise optical modules activations, and extraction of high confidence neutrino candidates. 
All three networks employ a transformer architecture that exploits inter-hit correlations through the attention mechanism. 
Applied sequentially, the pipeline achieves orders-of-magnitude speedup over the standard reconstruction chain.
Moreover, noise suppression neural network surpasses the accuracy of algorithmic noise suppression algorithms and provides estimate for time residuals of the signal hits, which is crucial for identification of track-like hits.
We address the domain shift between Monte Carlo simulations and experimental data by incorporating a domain adaptation technique, demonstrating improved agreement between the two domains. 
The resulting framework enables near-real-time event classification, with direct applications to multi-messenger alert systems and diffuse neutrino flux measurements.
}

\maketitle

%%%%%%%%%%%%%%%%%%%%
\section{Introduction}
%%%%%%%%%%%%%%%%%%%%
\label{sec:intro}

One of the central goals of astroparticle physics is the identification of sources of high-energy cosmic rays. Because cosmic rays are charged particles, their trajectories are deflected by galactic and extragalactic magnetic fields, and the observed arrival directions carry no direct information about the location of their origin. Neutrinos, by contrast, are electrically neutral and interact only weakly, traversing cosmological distances without deflection or absorption \cite{Gaisser:1995}. A detected high-energy neutrino therefore points back to its production site, providing a unique probe of the most violent astrophysical environments --- active galactic nuclei, gamma-ray bursts, and supernova remnants. This directional information makes neutrino astronomy a key pillar of multi-messenger astrophysics, in which observations across different messenger channels --- photons, gravitational waves, cosmic rays, and neutrinos --- are combined to build a coherent picture of transient and steady-state cosmic sources \cite{Bartos:2020, Aartsen:2018blazar}.

The detection of high-energy neutrinos exploits the Cherenkov effect. When a neutrino interacts with a nucleus in a transparent medium such as water or ice, the resulting charged secondary particles --- muons, electrons, or hadrons --- travel faster than the speed of light in the medium and hence emit Cherenkov radiation \cite{Tamm:1937}. Large-volume neutrino telescopes instrument cubic-kilometre-scale volumes with arrays of optical modules (OMs), each housing a photomultiplier tube sensitive to single Cherenkov photons. By recording the arrival times and amplitudes of the detected light across many OMs, the direction and energy of the parent neutrino can be reconstructed \cite{Halzen:2010}. Several neutrino telescopes are currently in operation or under construction, including IceCube at the South Pole \cite{Aartsen:2017jinst}, KM3NeT in the Mediterranean Sea \cite{Adrian-Martinez:2016}, and Baikal-GVD in Lake Baikal \cite{Avrorin:2019, Allakhverdyan:2021}, which is the focus of the present work.

A major challenge for all neutrino telescopes is the efficient processing of raw detector data. Noise activations of OMs --- caused primarily by natural bioluminescence and chemiluminescence of the medium --- can constitute up to 90\% of the hits within a triggered event and must be suppressed before event reconstruction can proceed. At the event level, the dominant background consists of extensive air showers (EAS): cosmic-ray interactions in the atmosphere produce bundles of muons that penetrate the water volume and generate Cherenkov light patterns similar to those of neutrino-induced muons \cite{Heck:1998}. The EAS-to-neutrino event ratio is of order $10^6$--$10^7$, and the standard data processing chain --- sequential application of noise filtering, track reconstruction, and directional cuts --- is computationally expensive. This constitutes a major bottleneck in the analysis workflow, particularly as the detector grows in size and the data volume increases.

In this paper, we propose to accelerate the Baikal-GVD data processing chain by introducing a set of dedicated neural networks. The pipeline consists of three stages. First, a classifier trained to distinguish neutrino events from EAS provides fast background suppression, eliminating 90\% EAS suppression while keeping 99\% of neutrino-induced events. Second, a noise filter for distinguishing between signal hits, that are due to propagation of relativistic particle in the medium, and noise hits. Third, a neutrino candidate extractor performs fine-grained event classification, identifying high-purity neutrino candidates suitable for astrophysical analyses and multi-messenger alerts. Applied sequentially, the three stages achieve orders-of-magnitude acceleration of the data analysis chain.

Each network is based on the transformer architecture \cite{Vaswani:2017}, whose self-attention mechanism naturally captures pairwise correlations between detector hits --- the relative space-time positions, charge ratios, and geometric consistency that encode the underlying physics. This allowed transformer neural networks to surpass the accuracy of algorithmic data analysis, which, while robust and interpretable, cannot take into account all complex correlations. The presented ML pipeline can be integrated with standard Baikal-GVD reconstruction: for example, improved noise suppression can improve the accuracy of the direction reconstruction algorithm.

A well-known limitation of machine learning in physics applications is the domain shift between Monte Carlo simulations and experimental data. In the context of Baikal-GVD, this shift may originate due to imperfect simulation of OM responses, imprecise water attenuation length, or simplifications of Monte Carlo simulations. If left unaddressed, these discrepancies cause the network to learn features that are specific to the simulated domain and do not generalise to data, make its predictions unreliable. We tackle this problem by incorporating a domain adaptation technique \cite{Ganin:2016} into the network training. The key idea is to penalise the network for learning domain-specific features, such as simulation artifacts, thus leading to physics-motivated feature extraction. We demonstrate that domain adaptation improves the agreement between MC-based predictions and experimental observations.

The paper is organised as follows. Section~\ref{sec:baikal} describes the Baikal-GVD detector, its event topologies, and the noise environment. Section~\ref{sec:mc} details the Monte Carlo simulations used for training and evaluation. Section~\ref{sec:nn} presents the neural network architecture and the three classification tasks. Section~\ref{sec:resuclts} reports the performance results, including the impact of domain adaptation. Section~\ref{sec:conclusion} summarises the findings and outlines future directions.

%%%%%%%%%%%%%%%%%%%%
\section{Baikal-GVD design}
%%%%%%%%%%%%%%%%%%%%
\label{sec:baikal}

Baikal-GVD (Gigaton Volume Detector) is a cubic-kilometre-scale neutrino telescope currently under construction in the southern basin of Lake Baikal, Russia. The detector targets high-energy astrophysical neutrinos in the TeV--PeV energy range \cite{Avrorin:2019, Allakhverdyan:2023diffuse}. Since the deployment of its first cluster in 2016, Baikal-GVD has been steadily expanded, reaching the volume of $0.8 $ m$^3$, and has already produced results on diffuse neutrino flux measurements \cite{Allakhverdyan:2023diffuse} and multi-messenger follow-ups \cite{Allakhverdyan:2021, Dik:2025}.
 
\begin{figure}[htbp]
 \centering
 \begin{subfigure}[b]{0.62\textwidth}
     \centering
     \includegraphics[width=\textwidth]{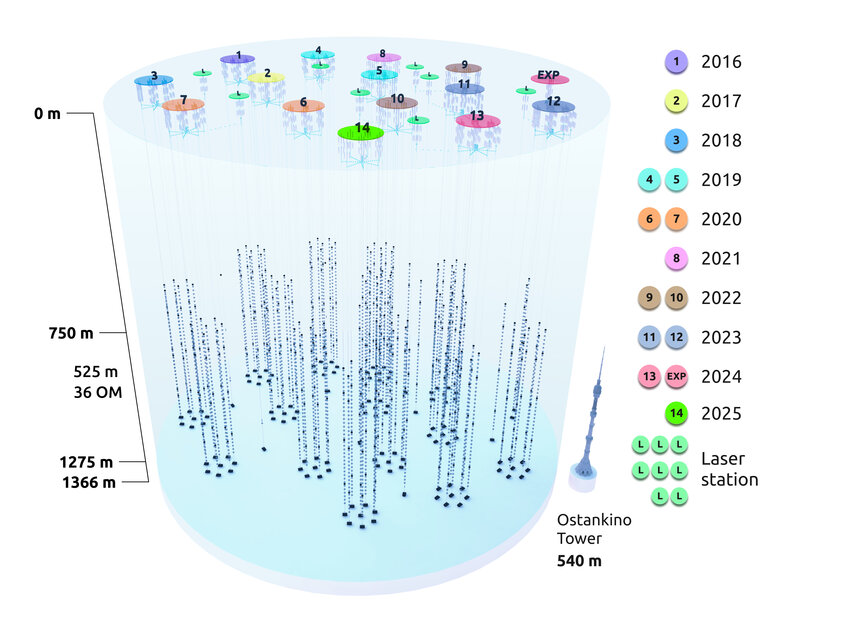}
     \caption{Baikal-GVD clusters layout.}
 \end{subfigure}
 \hfill
 \begin{subfigure}[b]{0.35\textwidth}
     \centering
     \includegraphics[width=\textwidth]{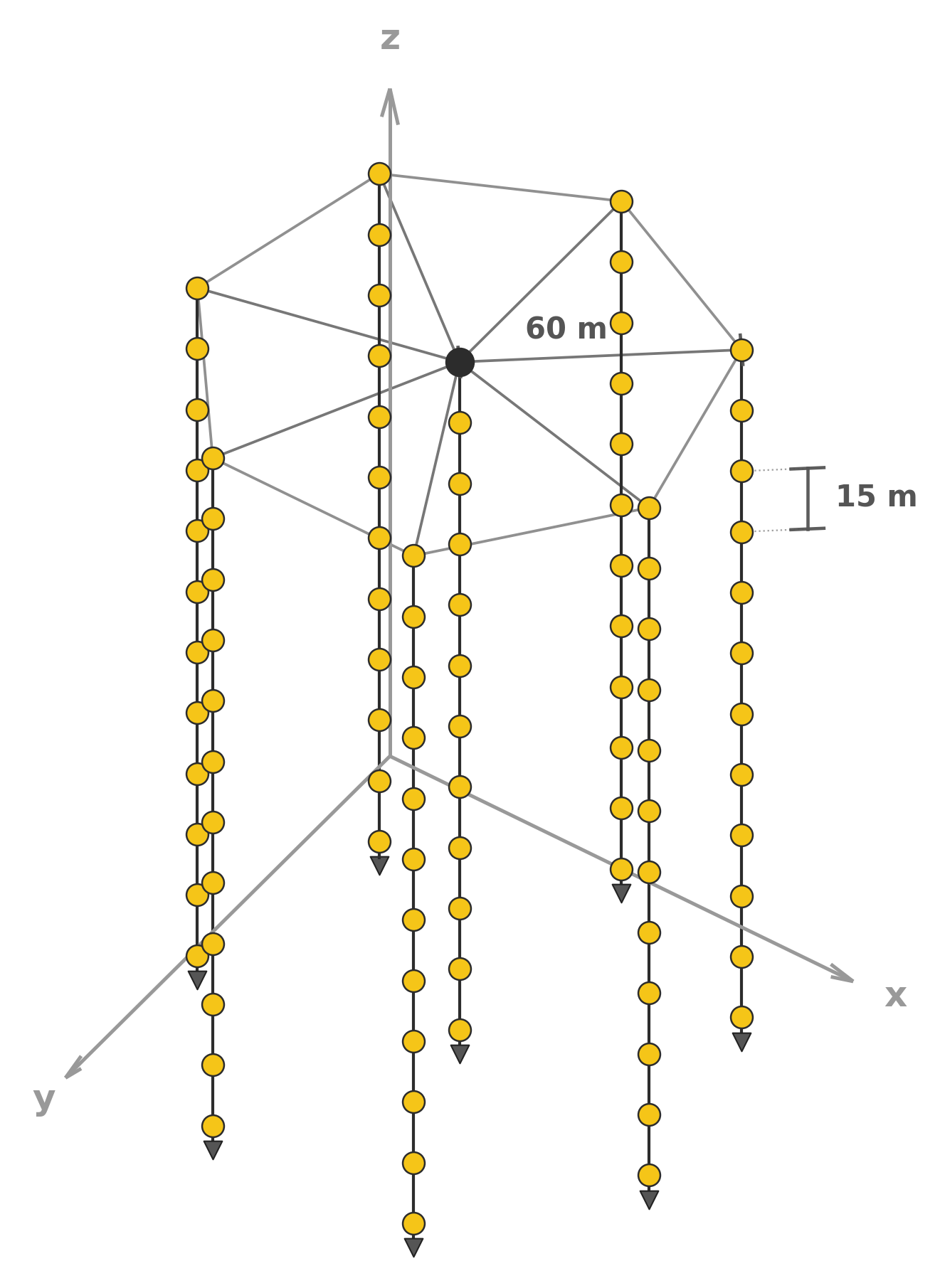}
     \caption{Single Baikal-GVD cluster.}
 \end{subfigure}
 \caption{Baikal-GVD design.}
 \label{baikal_detector}
\end{figure}

The detector has a modular architecture, illustrated in figure~\ref{baikal_detector}. The fundamental structural unit is the \textit{cluster}, with inter-cluster distance of 300 m \cite{Avrorin:2019, Safronov:2020}. Each cluster has eight vertical strings arranged in an approximate regular heptagon with one additional string at its center. The inter-string spacing within a cluster is approximately 60\,m. Each string carries 36 optical modules (OMs), oriented downward and spaced 15\,m apart along the vertical axis. Each OM houses a photomultiplier tube (PMT) sensitive to single photons, capable of recording both the arrival time and the integrated charge of each detected pulse.
 
\begin{figure}[htbp]
    \centering
    \includegraphics[width=0.75\textwidth]{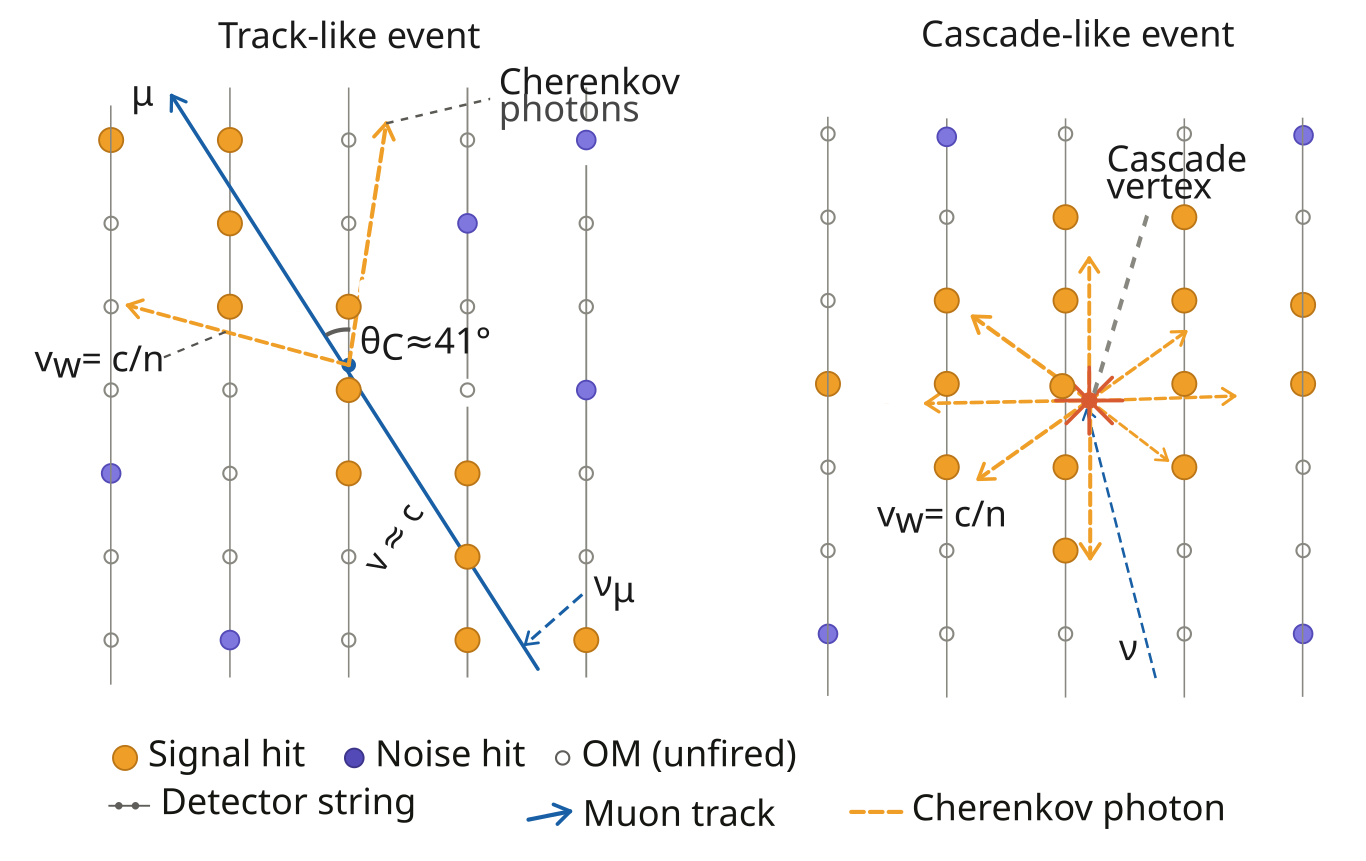}
    \caption{Illustration of neutrino-induced event topologies.}
    \label{event_topologies}
\end{figure}

Neutrino interactions in the instrumented water volume produce two principal event topologies, shown in figure~\ref{event_topologies}. Charged-current interactions of muon neutrinos generate relativistic muon tracks that traverse the detector over distances of up to several hundred metres, emitting Cherenkov radiation along their path. The Cherenkov photons propagate at a speed $v_w \approx c/1.35$ \cite{Avrorin:2002groupvelocity} and at a characteristic angle $\theta_C \approx 41^\circ$ relative to the muon direction. Interactions of all neutrino flavours via neutral currents, as well as charged-current interactions of electron and tau neutrinos, produce hadronic and electromagnetic cascades that emit Cherenkov light from a quasi-point-like source.
 
The dominant source of background at the single-hit level is natural water luminescence, primarily from bioluminescent organisms and chemiluminescent processes \cite{Avrorin:2019noise, Rjabov:2021}. This luminescence generates uncorrelated single-photon hits distributed across the detector at rates that vary seasonally and with depth, ranging from approximately 20 to 100\,kHz per OM \cite{Avrorin:2019noise, Dvornicky:2021}. Because these noise hits can constitute up to 90\% of the recorded data within a triggered event, effective noise suppression is a prerequisite for accurate event reconstruction \cite{Kharuk:2024}.
 
At the event level, the primary background consists of extensive air showers. Cosmic rays interacting in the atmosphere produce bundles of muons that penetrate the water and emit Cherenkov light along their tracks . Except for the case when neutrino interacts with medium inside the cluster, the key discriminating variable is the reconstructed arrival direction. Specifically, EAS muons arrive from above the horizon (downgoing), while astrophysical neutrinos are identified by their upgoing direction, having traversed the Earth. However, events arriving from directions near the horizon are difficult to classify, especially taking into account that Baikal-GVD clusters are 120 meters wide.
 
%%%%%%%%%%%%%%%%%%%%
\section{Monte Carlo simulations}
%%%%%%%%%%%%%%%%%%%%
\label{sec:mc}

In this paper, we focus on muon neutrinos. They are simulated in two physically distinct categories: atmospheric neutrinos and astrophysical neutrinos. Atmospheric neutrinos are produced in cosmic-ray interactions in the Earth's atmosphere and follow a steeply falling energy spectrum proportional to $ \sim E^{-3.7}$. Astrophysical neutrinos are generated with a softer spectrum, following $E^{-2}$, consistent with the measured diffuse flux \cite{Aartsen:2013science, Allakhverdyan:2023diffuse}. Neutrino interactions in the water are simulated, and the resulting secondary particles --- muons, electrons, and hadrons --- are propagated through the medium \cite{Safronov:2021EPJC}. The Cherenkov photons emitted along their trajectories are tracked using a dedicated Baikal-GVD photon propagation code that accounts for the optical properties of Lake Baikal water. The response of each OM, including the PMT quantum efficiency, angular acceptance, and electronic readout, is modelled in a separate detector simulation stage \cite{Avrorin:2016OM}. Absorption of neutrinos in the Earth is not included in the present simulation samples.
 
Uncorrelated noise hits are sampled independently at rates drawn from measured depth-dependent luminescence profiles and added to the simulated events \cite{Avrorin:2019noise, Dvornicky:2021}. This procedure ensures that the noise environment in the MC samples reflects the conditions encountered in experimental data.

The EAS background is simulated using the CORSIKA \cite{Heck:1998} and QGSJET-II-04 \cite{Ostapchenko:2006}, which model cosmic-ray interactions in the atmosphere and the subsequent development of particle cascades. The resulting muon bundles at the water surface are then propagated through the lake using the same photon propagation and detector response simulation as for the neutrino samples.

The triggering condition for identifying an \textit{event} is that adjacent OMs register signal above $Q_{\mathrm{th}}$ within a 100 ns time window \cite{Safronov:2021EPJC}. Depending on a season (noise conditions), the $Q_{\mathrm{th}}$ varies from 1 to 5 photo electrons. When this condition is fulfilled, the system records a 5~$\mu$s cluster data snapshot. Each OM activation (hit) is characterized by three quantities: its three-dimensional position $\mathbf{r}_i$, activation time, and the integrated charge $q_i$ deposited in the OM.

Several known limitations of the Monte Carlo simulations contribute to the domain shift between simulated and experimental data. First, the positions of the OMs in the deployed detector are subject to displacements due to water currents, which are not fully captured by the nominal geometry used in the simulation. Second, the modelling of individual OM efficiencies, including PMT gain variations and angular acceptance, is approximate. These discrepancies motivate the use of domain adaptation techniques, discussed in section~\ref{sec:nn}.

%%%%%%%%%%%%%%%%%%%%
\section{Neural network}
%%%%%%%%%%%%%%%%%%%%
\label{sec:nn}

%%%%%%%%%%%%%%%%%%%%
\subsection{Architecture}
%%%%%%%%%%%%%%%%%%%%
\label{sec:nn_arch}

The choice of neural network architecture is motivated by the structure of the data and the nature of the classification tasks. Each event consists of a variable-length set of hits, and the physical information relevant for classification is encoded in pairwise correlations between them --- their relative positions, time differences, and charge ratios. The transformer architecture is a natural fit for this problem: its self-attention mechanism computes pairwise attention scores between all input elements, allowing the network to learn which hit pairs carry the most discriminative information without imposing a fixed graph topology or ordering.

From a physical perspective, each transformer layer performs the following operation. For every OM hit, the self-attention mechanism evaluates its relation to all other hits in the event via learnable feature scalar products. The resulting score is used to compute a weighted combination of all OM representations, forming a message from all other hits to the given one. The feed-forward sublayer then updates the hit representation based on this aggregated information. After several such layers, each hit's representation encodes not only its own measured properties but also a summary of the global event context --- which other hits are spatially and temporally consistent with it, and how they are distributed across the detector. This progressively enriched, high-level representation can then be used to infer properties of individual hits (signal versus noise) or of the event as a whole (neutrino versus EAS).

Each hit is represented as an input ``token'' defined by a feature vector encoding its spatial coordinates $\mathbf{r}_i = (x_i, y_i, z_i)$, arrival time $t_i$, and charge $q_i$. These raw features are projected into a higher-dimensional embedding space through a learnable linear layer, increasing the representational capacity of the model. The embedded tokens are then processed by a stack of transformer encoder layers, each consisting of multi-head self-attention and feed-forward sublayers with residual connections and layer normalisation \cite{Vaswani:2017}. For hit-level classification, each output token is passed through a per-hit classification head that produces a scalar score. For event-level classification, the output tokens are aggregated into a single event-level representation via mean pooling and classification token technique \cite{Devlin:2019}, followed by an event-level classifier. We illustrate the neural network architecture in figure \ref{nn_arch} and provide neural network parameters in Appendix \ref{app:nn_details}.

\begin{figure}[htbp]
    \centering
    \includegraphics[width=0.95\textwidth]{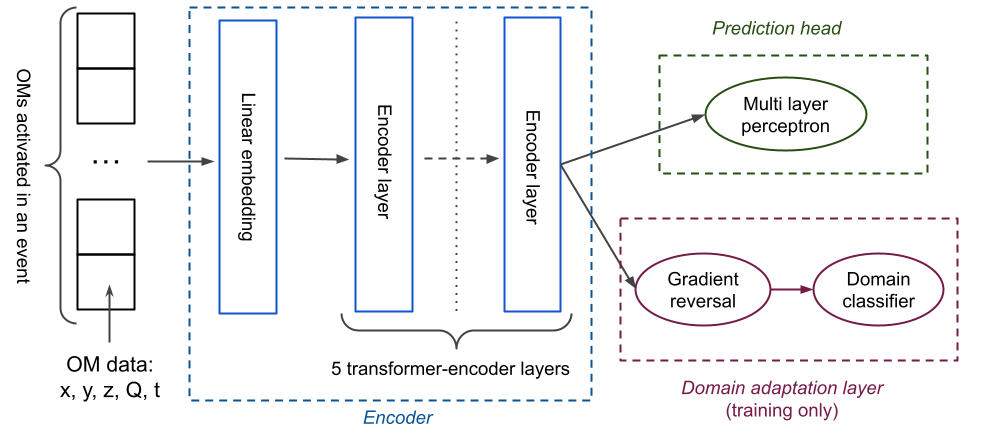}
    \caption{Neural network architecture. Prediction head is an multi layer perceptron for OM-wise (signal vs noise hits) or event-wise (neutrino vs EAS) predictions. Domain adaption layer forces the neural network to use domain-independent features.}
    \label{nn_arch}
\end{figure}

Several alternative architectures were considered. Convolutional neural networks (CNN) operate on data arranged in a 3D regular grid, which prohibits accounting for multiple activations of the same OM. While this can be resolved by introducing time slices, CNNs will struggle to identify temporal correlations between hits across different time slices. Recurrent neural networks and one-dimensional CNNs applied to the time-ordered sequence of hits offer a more natural data representation and can be effective for this type of problem. However, transformers are known to have superior representational capacity for modelling complex, non-local dependencies in sequential data. Graph neural networks provide a flexible framework for irregular point-cloud data, but their message-passing mechanism is local by construction: information between distant nodes must be relayed through multiple intermediate layers. This is known to result in the oversquashing, where long-range signals are progressively attenuated \cite{Alon:2021}. The transformer's self-attention mechanism avoids this limitation by computing direct pairwise interactions between all hits at every layer.

%%%%%%%%%%%%%%%%%%%%
\subsection{Domain adaptation}
%%%%%%%%%%%%%%%%%%%%
\label{sec:nn_da}

A fundamental challenge in deploying neural networks trained on Monte Carlo simulations is the domain shift between simulated and experimental data. As discussed in section~\ref{sec:mc}, this shift in Baikal-GVD arises from imperfect modelling of OM positions and efficiencies, and approximations in the water optical properties. A network that is not explicitly designed to be robust to these discrepancies may learn features that are specific to the simulated domain --- for instance, a particular geometry --- and fail to generalize to experimental conditions. This may result in uncontrolled systematic errors, compromising the enhanced reconstruction accuracy of the neural networks.

Domain adaptation (DA) comprises a class of techniques designed to render ML models robust against such systematic shifts between domains. The underlying principle is to encourage the network to extract features that are common to both domains, thereby ensuring that the learned representations capture fundamental physical properties rather than domain-specific artifacts. DA can be successfully applied in cases where patterns in the training data can be generalized to experimental data. For example, it was applied to generalize hand-written digit recognition from gray scale images to colored ones \cite{Ganin:2016}.  

To implement DA, the neural network is virtually partitioned into three components, as illustrated in figure \ref{nn_arch}: an \textit{encoder} (feature extractor), a \textit{prediction head}, and a \textit{domain adaptation layer} \cite{Ganin:2016}. The encoder, transformer-encoder in our case, maps input data from both domains into a shared latent feature space. The prediction head trained exclusively on labeled MC data, uses these features to perform the primary physics task of hit-level or event-level classification. Concurrently, the domain classifier, typically a shallow multilayer perceptron, attempts to identify the origin of the encoded features — whether they derive from MC simulations or experimental data.

The central element of this approach is a gradient reversal layer (GRL) positioned between the encoder and the domain classifier. During backpropagation, the GRL negates and scales by a factor $\lambda$ the gradient flowing from the domain classifier into the encoder. As a result, while the domain classifier is trained to maximize its ability to separate the two domains, the encoder receives the opposite learning signal and is driven to suppress exactly those features the domain classifier exploits. The result is an encoder that retains only features shared between MC and data, while discarding domain-specific patterns.

\begin{figure}[htbp]
    \centering
    \includegraphics[width=0.95\textwidth]{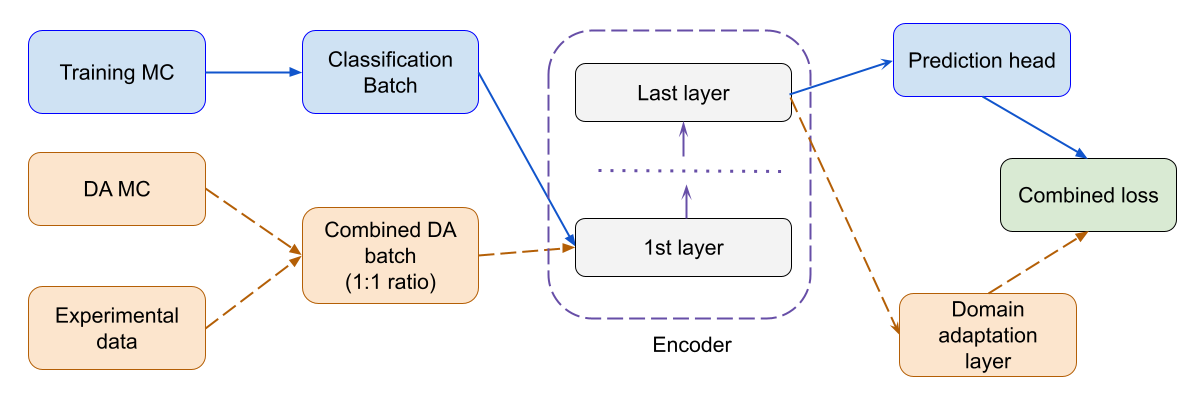}
    \caption{Data processing pipeline for neural network with domain adaptation. Blue and orange colors show the data flow for training and DA samples, correspondingly.}
    \label{da_baikal}
\end{figure}

The training procedure, in the general case, requires three datasets, as illustrated in figure \ref{da_baikal}. The \textit{training} Monte Carlo data is used to train the neural network using ground truth labels known from the simulation, following \textit{data} $\rightarrow$ \textit{encoder} $\rightarrow$ \textit{prediction head} path. The other two datasets are required for domain adaptation and follow \textit{data} $\rightarrow$ \textit{encoder} $\rightarrow$ \textit{domain adaptation layer} path. The first data set is the experimental data, which carries ``data'' label for the domain classifier. The last required data set is a \textit{DA} Monte Carlo sample for domain adaptation, providing artificial ``MC'' label for events. The reconstruction loss on MC events and the adversarial domain loss on both MC and experimental events are optimized jointly. At inference time, the domain adaptation module is inactive and introduces no additional computational overhead.

The third DA MC sample is required for unbiased domain adaptation. Specifically, if the distribution of some parameters differs significantly between the experimental and MC data, the domain classifier may interpret the mismatch as a domain-discriminating feature and, in trying to diminish it, introduce a bias. If training MC sample characteristics are close to the experimental data, it can be used instead of a separate DA MC sample.

For domain adaptation, we use experimental data and an equal mixture of EAS and atmospheric neutrino MC samples. Note that experimental data is dominated by \textit{downgoing} EAS events, making the zenith angle a domain discriminative feature. If this up-down asymmetry is not corrected, the gradient reversal would penalize the encoder for extracting features correlated with upgoing events, thus biasing the predictions. To resolve this, we randomly invert the $z$-axis with probability 0.5 for each event in both MC and experimental data samples during training.

We observe this effect in practice for the EAS suppression task (section~\ref{sec:fast_eas}), where the angular distribution asymmetry between MC and experimental data required careful selection and augmentation of experimental events to prevent the network from learning a spurious directional preference. When data sets are properly tuned, the domain classifier focuses on genuine detector modelling artifacts rather than on differences in the underlying physics distributions. 

%%%%%%%%%%%%%%%%%%%%
\section{Results}
%%%%%%%%%%%%%%%%%%%%
\label{sec:resuclts}

%%%%%%%%%%%%%%%%%%%%
\subsection{Preliminary EAS suppression}
%%%%%%%%%%%%%%%%%%%%
\label{sec:fast_eas}

The first stage of the pipeline aims to reject as many EAS as possible while retaining 99\% of neutrino events. At a background-to-signal ratio of order $10^5-10^7$, the achieved reduction of the event rate by approximately three orders of magnitude saves substantial processing time at the subsequent, more computationally expensive stages of the analysis chain.

For muon neutrino, the primary discriminating variable between EAS and neutrino events is the event arrival direction: neutrino events are upgoing, while EAS events are downgoing. The separation is, however, difficult for near-horizon events due to two factors. First, secondary cascades along the muon track can distort the apparent event direction. Second, since a single Baikal-GVD cluster spans approximately 120\,m in diameter, near-horizon tracks traverse only a short path within the instrumented volume, making it difficult to reliably estimate the arrival direction. For low-energy, single-cluster events this renders the two event classes hard to distinguish. Given that near-horizon events contribute the largest solid-angle exposure after integration over azimuth, reliable classification in this regime is essential.

\textbf{Neural network architecture and training.} 
To predict event type, EAS  versus neutrino, the output of the transformer-encoder is aggregated over OMs into a single event-level representation via the classification token technique \cite{Devlin:2019}. It is further passed through the event-level MLP classifier to produce a score $\xi$. Neutrino-like events are selected by applying a threshold on this score, allowing one to balance EAS rejection against neutrino efficiency. At the fast-suppression stage, the threshold is set to achieve high neutrino recall, leaving the identification of neutrino events to later stages of data analysis.

Experimental data contains many low‑quality events with very few signal hits, which are not suitable for further analysis. To reject such events as well, events with fewer than 6 signal hits (according to MC truth) are marked as background and the network is trained to reject them alongside EAS. 

The network is trained using the focal loss \cite{lin2017focal}. The training MC dataset includes EAS, atmospheric, and cosmogenic neutrinos in a ratio of 2:1:1. This ensures that neural networks sees all types of events during training, which reduces event-selection bias.

For domain adaptation, we use experimental data and an equal mixture of EAS and atmospheric neutrino MC samples. Note that experimental data is dominated by \textit{downgoing} EAS events, making the zenith angle a domain discriminative feature. If this up-down asymmetry is not corrected, the domain adaptation layer would penalize the encoder for extracting features correlated with upgoing events, thus biasing the predictions. To resolve this, we randomly invert the $z$-axis with probability 0.5 for each event in both MC and experimental data samples during training.

\begin{figure}[htbp]
\centering
\begin{minipage}{.48\textwidth}
  \centering
  \includegraphics[width=.9\linewidth]{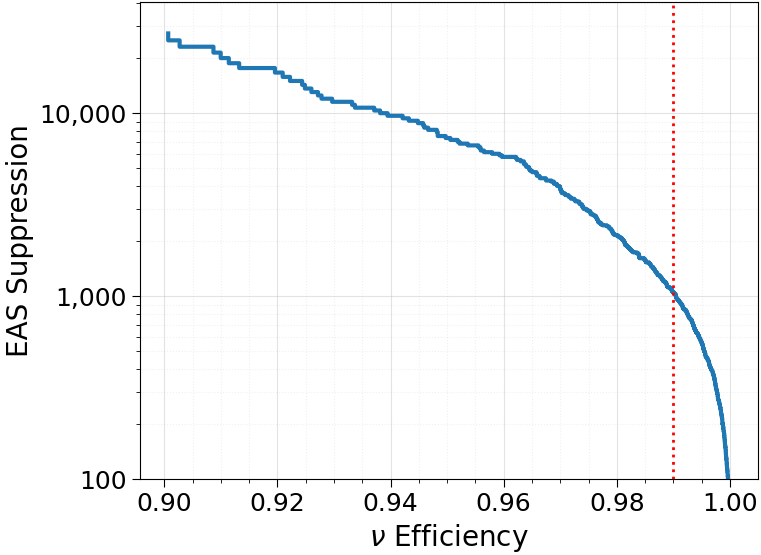}
  \captionof{figure}{EAS suppression factor vs. $\nu_{\mu}$ detection 
    efficiency for the preliminary background suppression network.
    The $\nu^{atm}$ and $\nu^{cosmo}$ neutrino samples include
    $ \approx 100,000$ events each), tested against 
    $N_\mathrm{EAS} \approx 600,000$ EAS-induced events.
    The red dotted line represents the $99\%$ neutrino recall level.
}
  \label{prefilter_metrics}
\end{minipage}%
\hfill
\begin{minipage}{.48\textwidth}
  \centering
  \includegraphics[width=.9\linewidth]{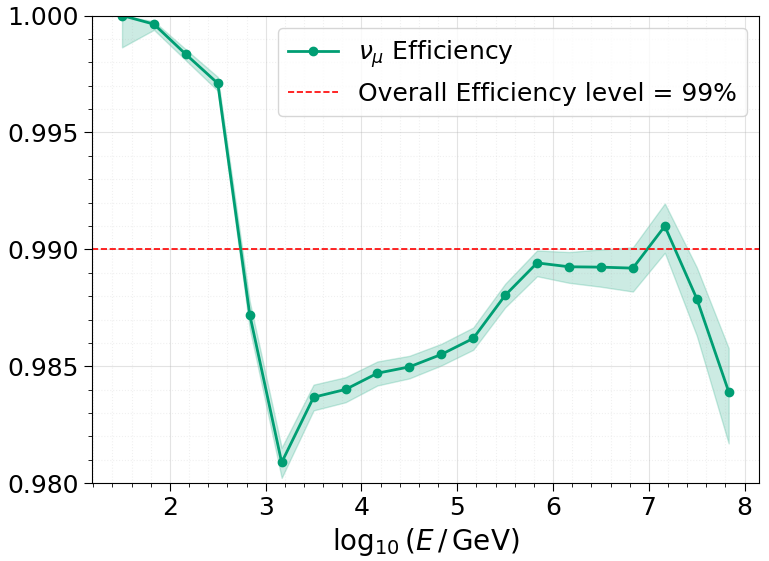}
  \captionof{figure}{MC $\nu_\mu$ detection efficiency as a function of the primary particle energy for events passing standard cuts. The performance metrics are computed $\xi$ corresponding to 99\% overall $\nu_\mu$ selection efficiency. The shaded areas represent the interquartile range for each energy bin.}
  \label{prefilter_metrics_vs_e}
\end{minipage}
\end{figure}

\textbf{Results.}
On MC data with at least 8 singla hits on at least 2 string the event classifier achieves high EAS rejection rates while retaining the majority of neutrino events, Figure \ref{prefilter_metrics}. At a threshold $\xi=0.761$ tuned for fast suppression, the neutrino recall is 99\% with an EAS suppression factor of 992, promising a data analysis speedup by the same factor. In what follows, we refer to event selection cuts $n_{\text{hit}} \geq 8 $ and $n_{\text{str}} \geq 2$ as \textit{standard cuts}.

Figure \ref{prefilter_metrics_vs_e} depicts the dependence of the neutrino recall on primary particle energy. At low energies the muon leaves a clear track, which allows for effective discrimination between EAS- and neutrino-induced events. At higher energies the muon enters the stochastic energy‑loss regime, which makes the two types of events hard to distinguish due to scattered signal hits. Finally, at very high energies cosmogenic neutrino yield higher brightness than EAS, which serves as a powerful discriminating feature. The high fraction of $\nu^{cosm}_\mu$ used for training neural network thus introduces a bias into the classification. We believe, however, that instead of minimizing this bias it is better to keep all high energy events, both EAS- and $\nu^{cosm}_\mu$-induced, since they deserve special attention and can be filtered out at later stages of data analyzes. 

\begin{figure}[htbp]
    \centering
    \includegraphics[width=0.8\textwidth]{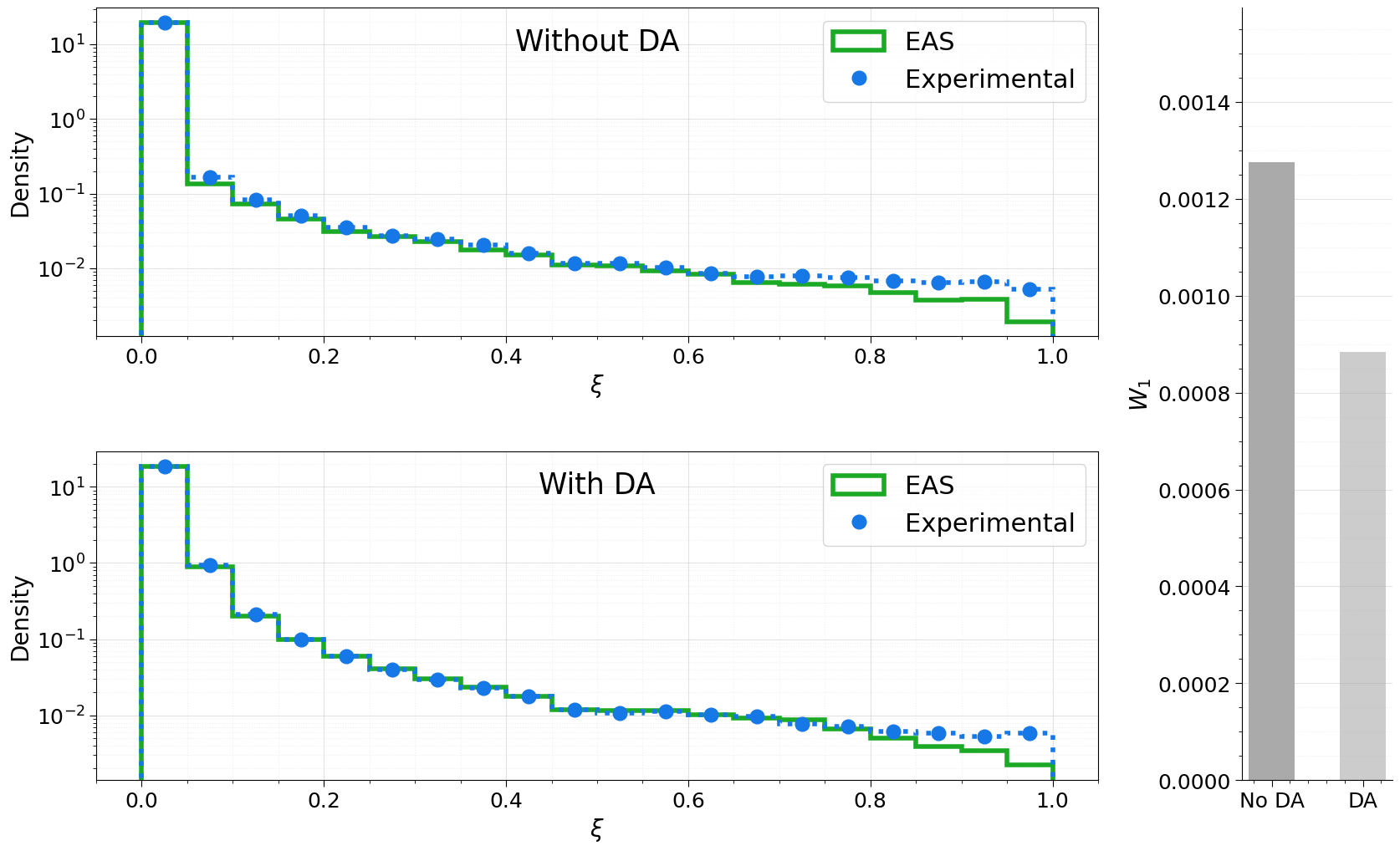}
        \caption{
    Distributions of the classifier output $\xi$ for MC simulated EAS and experimental events
    after reconstruction-quality cuts, without domain adaptation (top) and with domain adaptation (bottom).
    The Wasserstein distance $W_1$ between the two distributions is shown in the bar chart on the right.
    }
    \label{prefilter_distrs_reco_cuts}
\end{figure}

\textbf{Validation on experimental data.}
To verify that the neural network is robust with respect to the MC–data domain shift, we compared distributions of the classifier output on experimental and MC EAS events. Since EAS constitute the majority of experimentally observed events, these two diagrams should be close. Unlike MC, experimental data contains a large fraction of events with no signal hits, triggered purely by water luminescence. To select data subsets for proper comparison, we apply the following set of reconstruction-quality cuts using the standard track reconstruction algorithm:
\begin{itemize}\setlength{\itemsep}{0pt}
    \item number of signal hits $n_\text{hits} \geq 8$ and number of signal strings $n_\text{strings} \geq 3$;
    \item covariance matrix convergence status equal to 3 (fully successful fit);
    \item log-likelihood of the hit pattern $\log_{10} p_\text{hit} > -9$;
    \item normalized fit quality $\log_{10}\!\left(F / (n_\text{hits} - 5)\right) < 2$, where $F$ is the minimizer objective value;
    \item fraction of hit triplets $n_\text{triplets} / n_\text{hits} > 0.1$ (track-topology requirement);
    \item reconstructed zenith angle uncertainty $\sigma_\theta < 2.5$~rad;
    \item vertical track length $z_\text{dist} > 70$~m;
    \item $z$-coordinate of the event center $z_\text{center} < 220$~m;
    \item number of minimizer calls $< 350$ (convergence quality);
    \item reconstructed zenith angle $\theta_\text{rec} \in (80^\circ,\,180^\circ)$.
\end{itemize}
With the imposed cuts, the classifier output distributions for MC EAS and experimental data are close, Figure \ref{prefilter_distrs_reco_cuts}. This proves that neural network prediction are reliable. We also noticed that domain adaptation reduces the Wasserstein distance between the two distribution from 0.0013 to 0.0009. Based on this improvement we decided to use the neural network with implemented domain adaptation technique.

To verify that neural network does not drop neutrino events, we considered its output on the known neutrino candidates from the standard event reconstruction pipeline. On the experimental sample with 125 neutrino events, all of them have $\xi > 0.85$, which is well above the EAS-suppression threshold.

%%%%%%%%%%%%%%%%%%%%
\subsection{OM noise suppression}
%%%%%%%%%%%%%%%%%%%%
\label{sec:noise}

The hit-level noise filter is applied to events surviving the EAS suppression stage and is designed to improve the quality of the hit sample before further processing. Its purpose is to separate true signal hits --- photons originating due to radiation of a relativistic particle --- from noise hits caused by stochastic water luminescence. The exatracted signal hits will be used by all downstream event reconstruction algorithms, including neural-network-based.

The underlying physical principle that makes hit-level classification possible is the distinction between the correlated spatial and temporal structure of signal hits and the random, uncorrelated noise hits. Signal hits originate from Cherenkov photons emitted by a relativistic particle traversing the detector; they are therefore constrained by the geometry of Cherenkov light propagation and form characteristic patterns. Noise hits, by contrast, arise from stochastic luminescence processes and are distributed randomly across the detector in both space and time. The self-attention mechanism of the transformer is well suited to exploit this contrast: by evaluating all pairwise hit relations simultaneously, the network identifies groups of mutually consistent hits that conform to a physical propagation pattern.

\textbf{Neural network architecture and training.} 
Each output OM token produced by the transformer encoder is passed through a per-hit MLP classification head yielding a scalar score $\xi \in [0,1]$, interpreted as the probability that the hit belongs to the signal class. Signal hits are then selected by applying a threshold on this score, which allows one to trade precision for recall according to the requirements of the downstream analysis.

In addition to signal/noise classification, the network simultaneously predicts the \textit{time residual} $t_\text{res}$ of each signal hit, defined as the difference between the observed photon arrival time and the arrival time expected under the Cherenkov propagation model for muon track. A small positive $t_\text{res}$ is characteristic of direct, unscattered Cherenkov photons (track-like hits), while large values arise from scattered photons or from secondary electromagnetic cascades along the track (cascade-like hits). This auxiliary prediction allows downstream reconstruction
algorithms to select only track-like hits when required, or to use the estimated $\hat{t}_\text{res}$ as a continuous input feature in likelihood-based methods.

The total training objective combines three loss terms:
\begin{equation}
\mathcal{L} = \mathcal{L}_{\text{cls}} + k_1 \, \mathcal{L}_{t_{\text{res}}} + k_2 \, \mathcal{L}_{\text{DA}},
\label{eq:total_loss}
\end{equation}
where $\mathcal{L}_{\text{cls}}$ is the binary cross-entropy loss for signal/noise classification, $\mathcal{L}_{t_{\text{res}}}$ is the time-residual regression loss computed exclusively on signal hits, and $\mathcal{L}_{\text{DA}}$ is the adversarial domain-adaptation loss described in Section~\ref{sec:nn_da}. The coefficients $k_1$ and $k_2$ control the relative contribution of the auxiliary objectives. We set $k_1 = 0.5$ and $k_2 = 10^{-3}$, as they provide optimal metrics on the valideation data set. 

The time-residual loss uses a piecewise formulation that transitions from mean absolute error to a logarithmic regime at a threshold $t^{*} = 15\,$ns:
\begin{equation}
\mathcal{L}_{t_{\text{res}}} = \frac{1}{N_{\text{sig}}} \sum_{i \in \text{sig}} \ell\!\left(|t_{\text{res},i}|,\; \hat{t}_{\text{res},i}\right), \qquad
\ell(t, \hat{t}) =
\begin{cases}
|t - \hat{t}|, & \text{if } t < t^{*}, \\[4pt]
\bigl|\log_{10} t - \log_{10} \hat{t}\bigr|, & \text{if } t \geq t^{*},
\end{cases}
\label{eq:tres_loss}
\end{equation}
where the sum is taken over all signal hits in the batch and $N_{\text{sig}}$ is the total number of signal hits in a batch. The logarithmic scaling for large residuals softens the loss for heavily scattered photons, preventing those hits from dominating the gradient and allowing the network to achieve high accuracy in the physically critical regime of small residuals.

A five-layer transformer-encoder form a shared feature backbone common to all three tasks. The classification branch applies a linear projection to the per-hit representations, producing two logits per hit. The time-residual regression branch uses a two-layer MLP with 256 hidden units and GELU activation, producing a single per-hit estimate of $|t_{\text{res}}|$.

The network is trained on a mixture of EAS, atmospheric neutrino, and cosmogenic neutrino MC events in the ratio 2:1:1, ensuring balanced representation of all types of events. For domain adaptation we use the same sample of MC events mixed with equal number of experimental data in each batch. We also keep random $z$-axis flips active to neglect the pressure from domain classifier on encoder for extracting direction-related event properties.

\begin{figure}[htbp]
  \centering
  \includegraphics[width=0.95\textwidth]{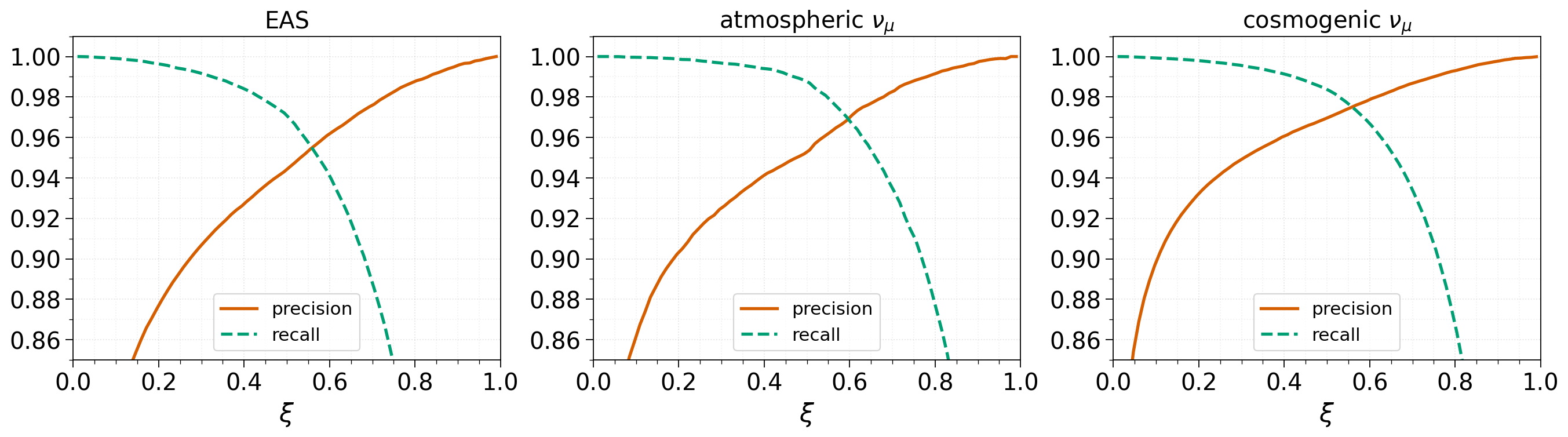}
  \caption{Hit-level precision and recall as a function of the classification threshold~$\xi$ for EAS, atmospheric and cosmogenic neutrino events after applying standard cuts.}
  \label{fig:pr_vs_threshold}
\end{figure}

\textbf{Results.}
We evaluate the noise filter performance using hit-level precision and recall as a function of the classification threshold~$\xi$, Figure~\ref{fig:pr_vs_threshold}. After imposing event selection cuts $n_{\text{hit}} \geq 8$ and $n_{\text{str}} \geq 2$, both precision and recall exceed 0.95 at $\xi = 0.5$ for all event types, surpassing the accuracy of the standard scan-fit noise filtering algorithm~\cite{allakhverdyan2021efficient}. 

\begin{figure}[htbp]
  \centering
  \begin{subfigure}[b]{0.95\textwidth}
    \includegraphics[width=\linewidth]{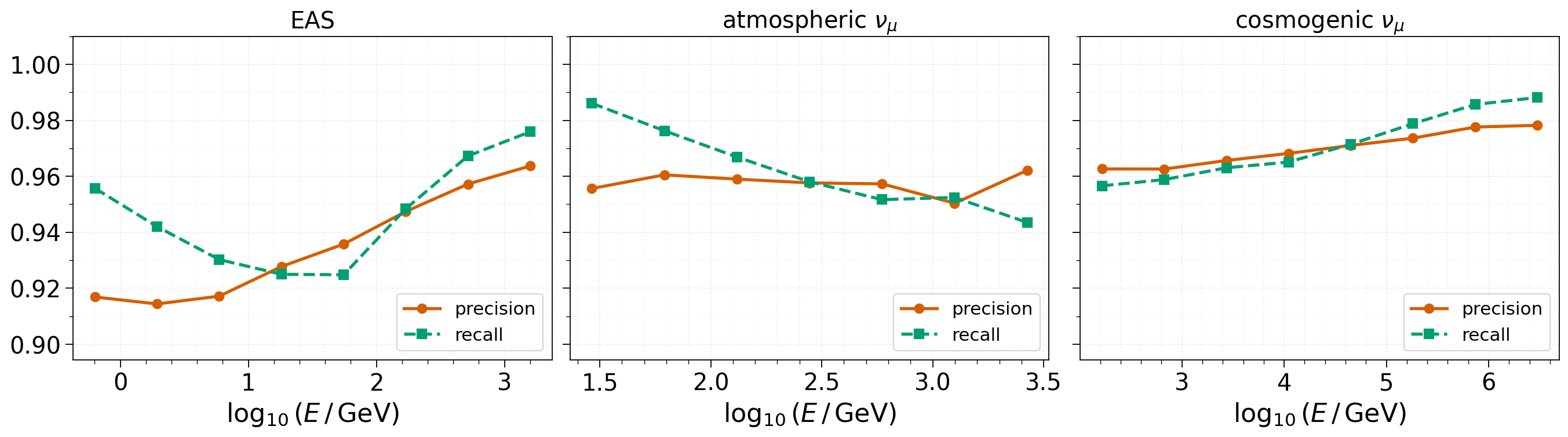}
  \end{subfigure}
  
  \begin{subfigure}[b]{0.95\textwidth}
    \includegraphics[width=\linewidth]{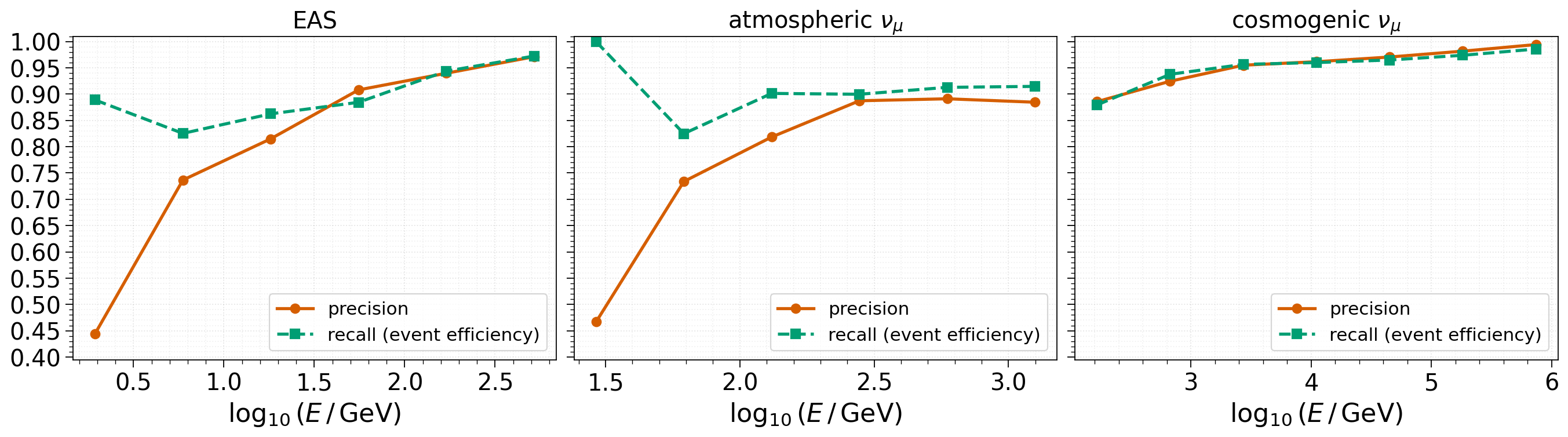}
  \end{subfigure}
  
  \caption{Precision and recall metrics for signal hits selection (upper panel) and event selection (lower panel) on EAS, atmospheric and cosmogenic neutrino events as function of primary particle energy. We impose standard cuts according to neural network predictions with classification threshold $\xi = 0.5$.}
  \label{om_ev_prec_recall}
\end{figure}

Figure~\ref{om_ev_prec_recall} shows the energy dependence of the same metrics. The network maintains high event-level recall (fraction of events passing the selection) across the full energy range, with performance improving at higher energies where the signal-to-noise ratio is more favourable. At low energies, the event-level precision decreases, particularly for EAS and atmospheric neutrino events. This is expected: low-energy particles produce fewer Cherenkov photons, so events near the selection threshold ($n_{\text{hit}} \approx 8$) require nearly all of their hits to be classified as signal, inevitably including some misidentified noise hits. The hit-level precision (upper panel) remains high throughout, confirming that the drop in event-level precision is an artefact of the selection threshold applied to low-multiplicity events, not a failure of the per-hit classifier.

\begin{figure}[htbp]
\centering
\begin{minipage}{.48\textwidth}
  \centering
  \includegraphics[width=.9\linewidth]{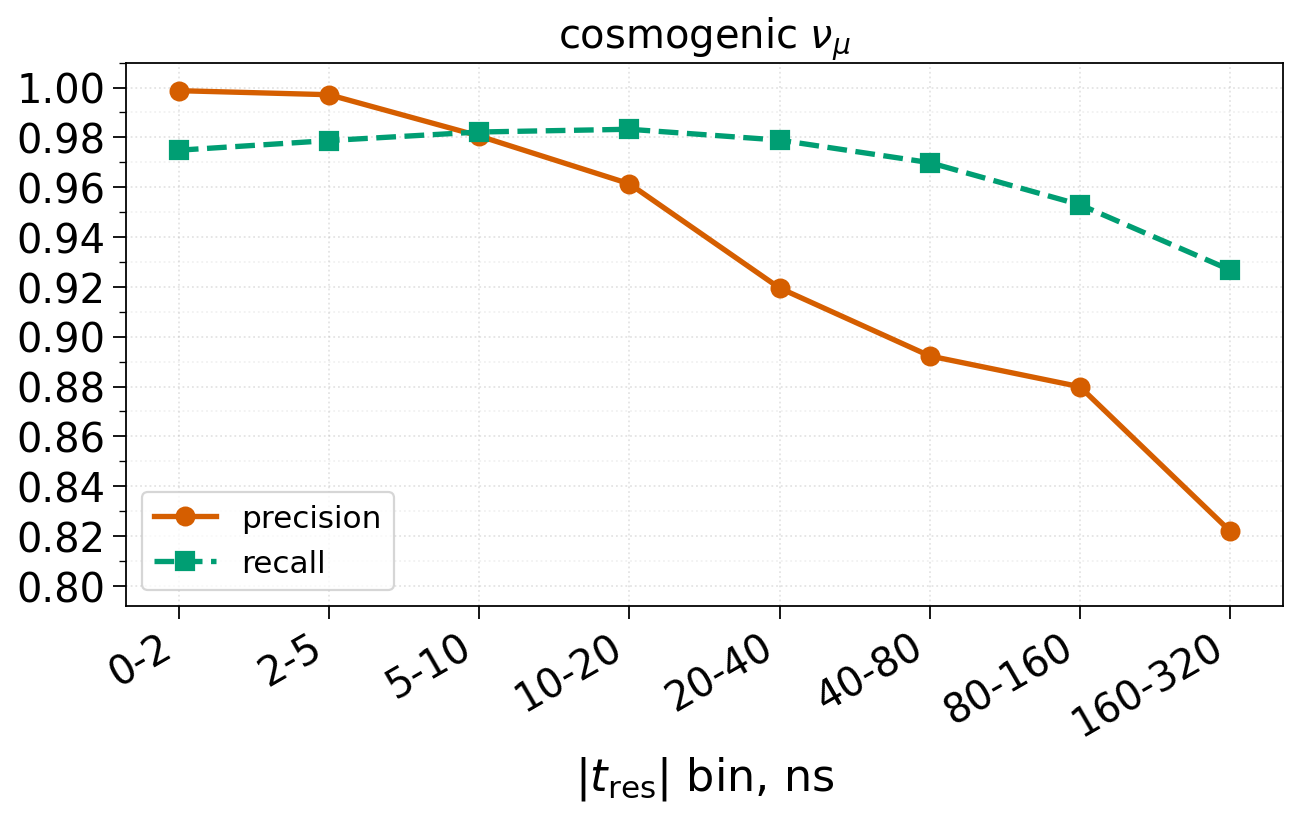}
  \captionof{figure}{Hit-level precision and recall for cosmogenic neutrino 
  events as a function of the true $|t_{\text{res}}|$ at classification 
  threshold $\xi = 0.5$.
}
  \label{fig:tres_dep}
\end{minipage}%
\hfill
\begin{minipage}{.48\textwidth}
  \centering
  \includegraphics[width=.9\linewidth]{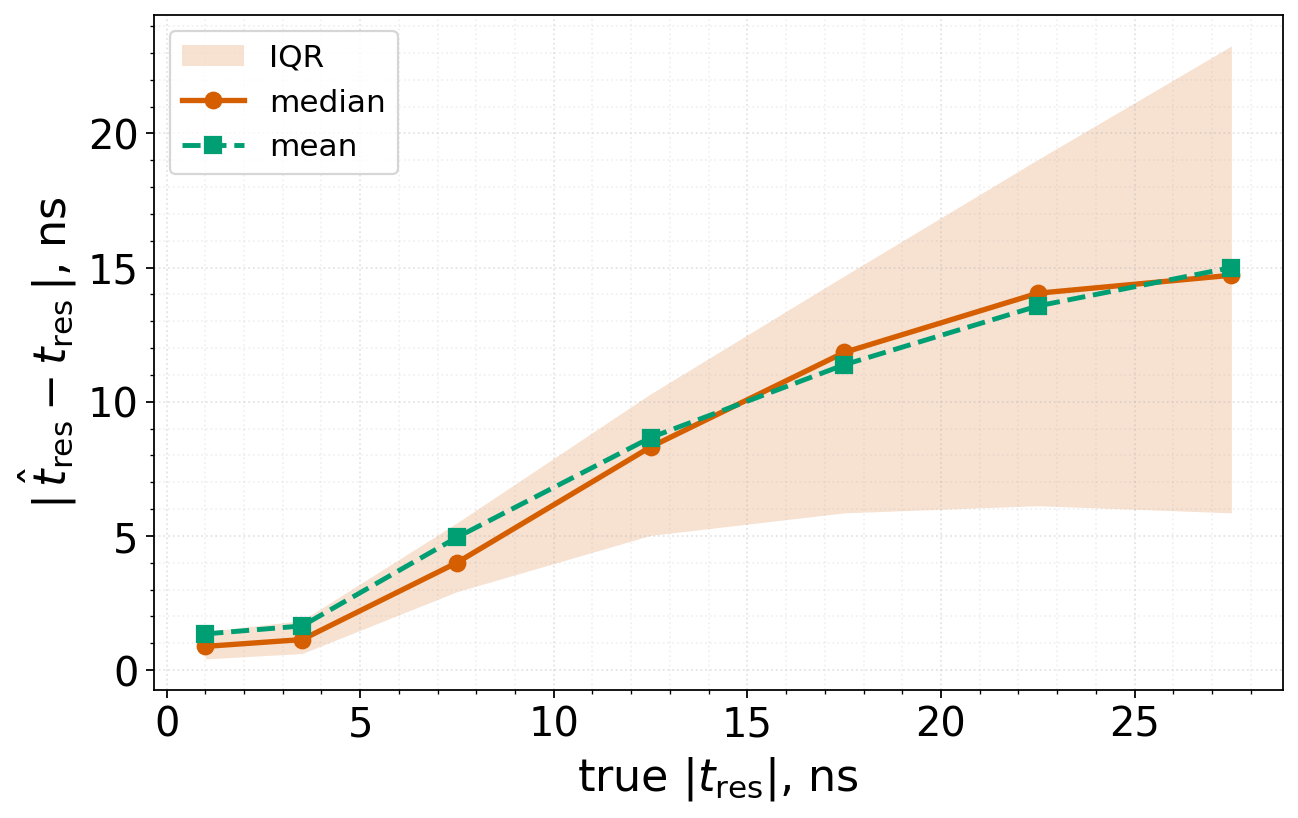}
  \captionof{figure}{Time-residual prediction error 
  $|\hat{t}_{\text{res}} - t_{\text{res}}|$ as a function of the true $|t_{\text{res}}|$ for signal hits. The shaded band depicts the interquartile range.}
  \label{fig:tres_prediction}
\end{minipage}
\end{figure}

Figure~\ref{fig:tres_dep} shows that the classifier maintains high recall across the full range of time residuals, including the large-$|t_\text{res}|$ regime corresponding to cascade-like, scattered photons. This ensures that cascade-induced hits are not preferentially discarded, which would otherwise bias the reconstructed event topology.

The network yields accurate estimates of the hit time residuals, Figure~\ref{fig:tres_prediction}, with a median prediction error below 2\,ns for hits with $|t_{\text{res}}| < 5\,$ns. This enables selecting track-like hits by imposing a cut on the estimated $\hat{t}_{\text{res}}$ value, which is important for track-like event reconstruction algorithms that require a clean sample of direct Cherenkov hits.

\begin{figure}[htbp]
  \centering
  \includegraphics[width=0.95\textwidth]{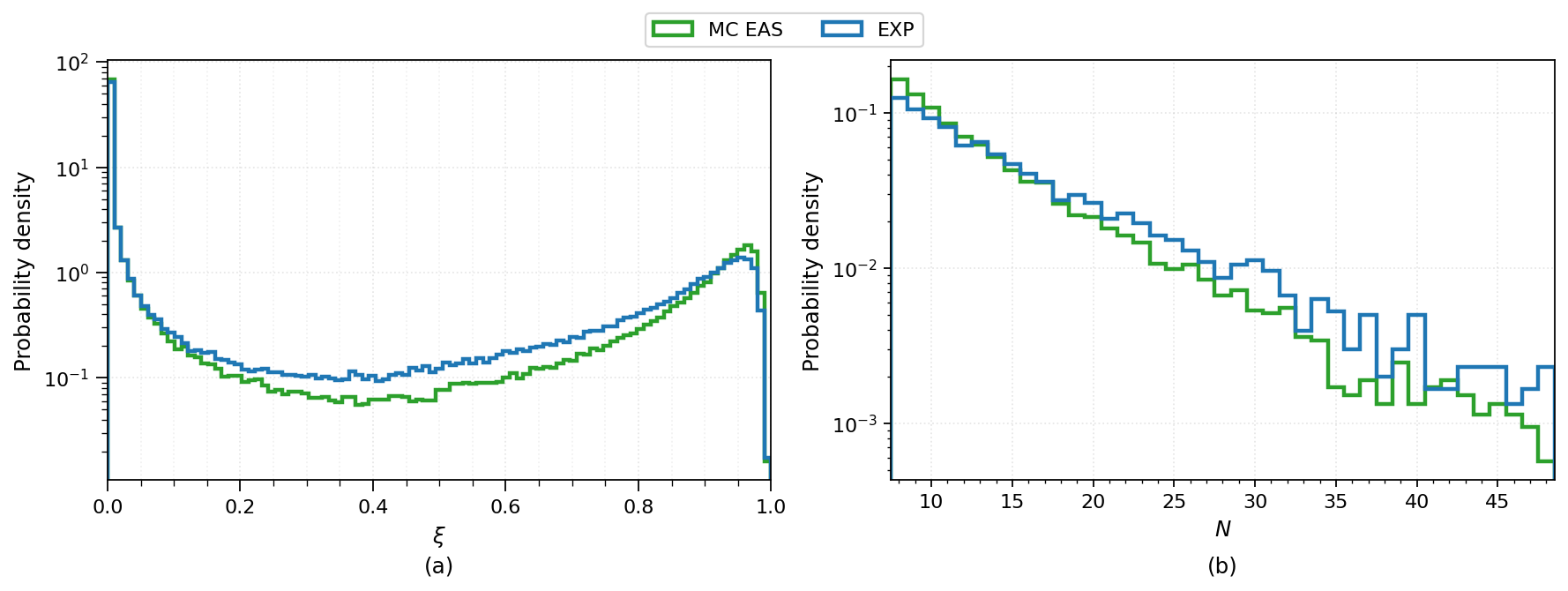}
  \caption{Distribution of the neural network output score (a) and signal hit multiplicity (b) for MC EAS and experimental data after the nn-based standard cuts with the classification threshold of $\xi = 0.5$.}
  \label{mc_exp_compare_hits}
\end{figure}

\textbf{Validation on experimental data.}
To assess the reliability of the noise filter on experimental data, we select experimental events for which the network identifies at least 8 signal hits on at least 2 distinct strings, which we dub \textit{nn-based standard cuts}. This requirement removes noise-triggered events that have no analogue in the MC simulation, enabling a direct comparison of network outputs on EAS MC and experimental
data. Figure~\ref{mc_exp_compare_hits} shows that the score distribution and the predicted signal-hit multiplicity are in close agreement between MC and data, confirming that the classifier generalizes reliably to experimental conditions. Residual differences are attributed to known MC imperfections, including imprecise modelling of OM quantum efficiency and light scattering.

\begin{figure}[htbp]
  \centering
  \includegraphics[width=0.95\textwidth]{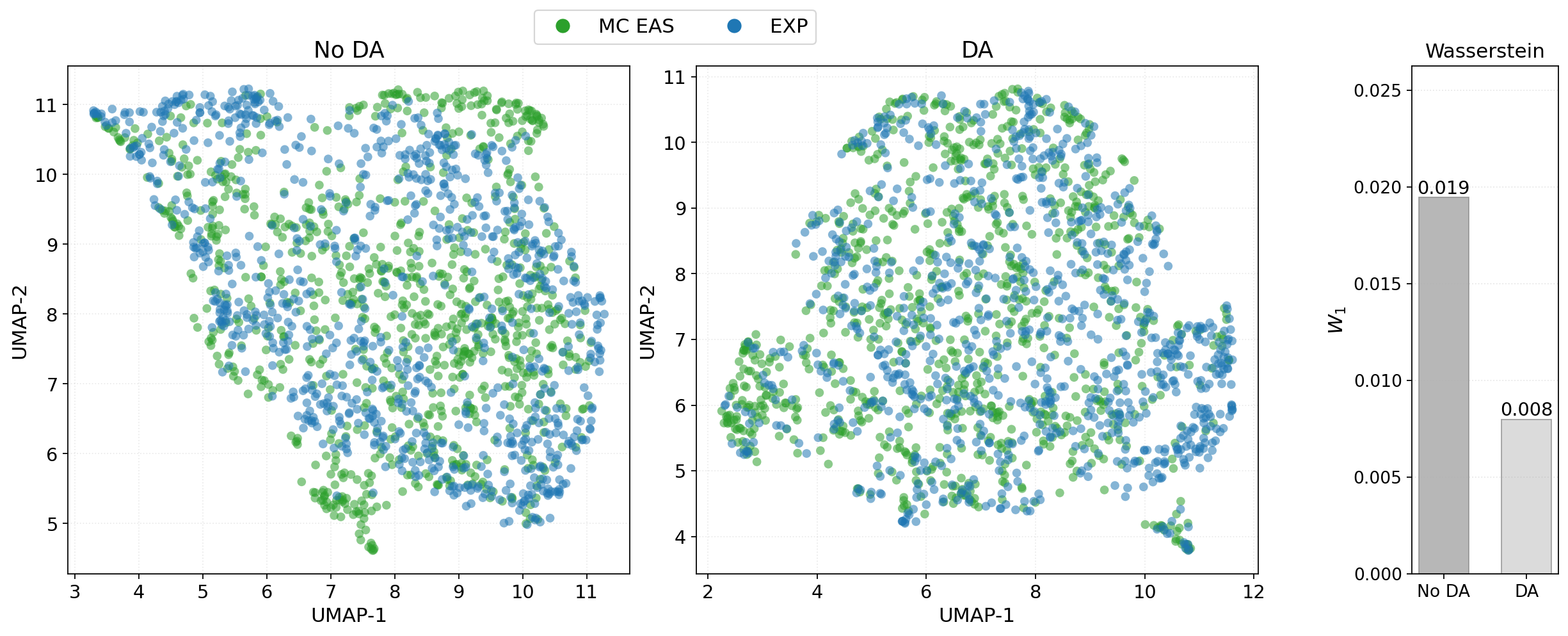}
  \caption{UMAP projection of event-level latent representations from the last transformer-encoder layer for MC EAS and experimental data after the nn-based standard cut. Left: network trained without domain adaptation. Centre: network trained with domain adaptation. Right: Wasserstein distance $W_1$ between MC EAS and experimental hit-level score distributions.}
  \label{fig:umap_da}
\end{figure}

To examine domain adaptation at the representation level, we extracted event-level latent vectors from the last transformer-encoder layer and projected them onto two dimensions using UMAP~\cite{McInnes2018}. As Figure~\ref{fig:umap_da} demonstrates, the MC and experimental clusters are more strongly overlapping when the network is trained with DA, indicating that the encoder has discarded simulation-specific features in favour of physics-driven representations shared by both domains. To quantify the effect, we evaluated the Wasserstein distance between the MC EAS and experimental hit-level score distributions: DA reduces the discrepancy from $0.019$ to $0.008$.

%%%%%%%%%%%%%%%%%%%%
\subsection{Neutrino candidates identification}
%%%%%%%%%%%%%%%%%%%%
\label{sec:nu_candidates}

At the final stage, a second event-level classifier is applied to identify neutrino-induced events within the EAS-suppressed, noise-filtered sample. Prior noise suppression improves the quality of the input to this classifier: with noise hits removed, the remaining hit pattern more clearly reflects the underlying event topology, enabling the network to achieve better discrimination between neutrino and EAS events. The main difficulty at this stage is near-horizon
events, where the arrival directions of EAS muons and neutrino-induced muons are nearly indistinguishable and where the limited chord length of the track within a single cluster provides little geometric leverage. The classifier must therefore achieve high purity with minimal loss of neutrino efficiency in this challenging angular regime, while remaining
robust against the MC--data domain shift. From the utility perspective, such neural network would allow obtaining a clean sample of neutrino candidates suitable for astrophysical analyzes and real-time multi-messenger alert generation. 

\textbf{Neural network architecture and training.}
The MC training dataset was constructed from events pre-processed by the noise suppression model described in Section~\ref{sec:noise}. Only hits with a neural network signal-probability score exceeding $0.8$ are retained. This threshold is stricter than the $\xi = 0.5$ used to evaluate the noise filter: the tighter cut provides a cleaner hit sample as input, accepting a modest reduction in signal-hit recall in exchange for a substantially lower noise contamination. This also allows to reduce the effect of imperfect MC simulations, which were use for training noise suppression neural network. Events with fewer than 5 surviving hits after this cut are discarded as insufficiently reconstructable. The resulting dataset contains approximately $5{,}000{,}000$ events in total, with EAS, atmospheric, and cosmogenic neutrino events in the ratio 2:1:1 --- corresponding to an equal split between the EAS class and the combined neutrino class.

The network is trained using the focal loss~\cite{lin2017focal}, which suppresses the gradient contribution from well-classified examples and concentrates learning on hard, near-threshold events. This is particularly beneficial for the near-horizon regime, where the decision boundary is most ambiguous. Domain adaptation is applied using a linearly scheduled adversarial penalty, and random $z$-axis flips are retained to prevent the domain classifier from exploiting the up-down asymmetry of the experimental data (see Section~\ref{sec:nn_da}). Data augmentation further includes Gaussian noise applied independently to each of the five input features ($x$, $y$, $z$, $t$, $q$) to mimic MC modelling uncertainties, and random azimuthal rotation of hit coordinates, which exploits the approximate azimuthal symmetry
of the detector and effectively multiplies the training set size.

\begin{figure}[htbp]
    \centering
    \subcaptionbox{\label{fig:nu_classifier_SuppEff}}{%
        \includegraphics[width=0.32\textwidth]{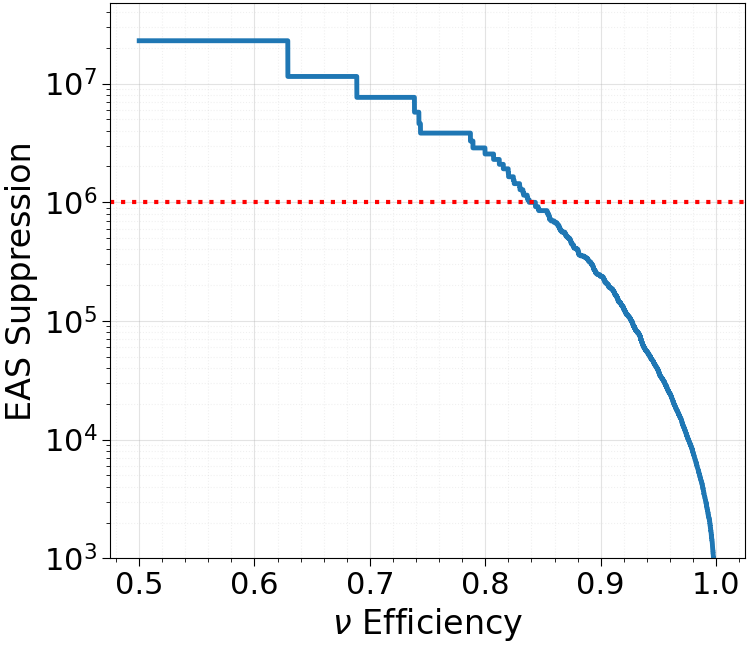}
    }\hfill
    \subcaptionbox{\label{fig:nu_classifier_eff_theta}}{%
        \includegraphics[width=0.32\textwidth]{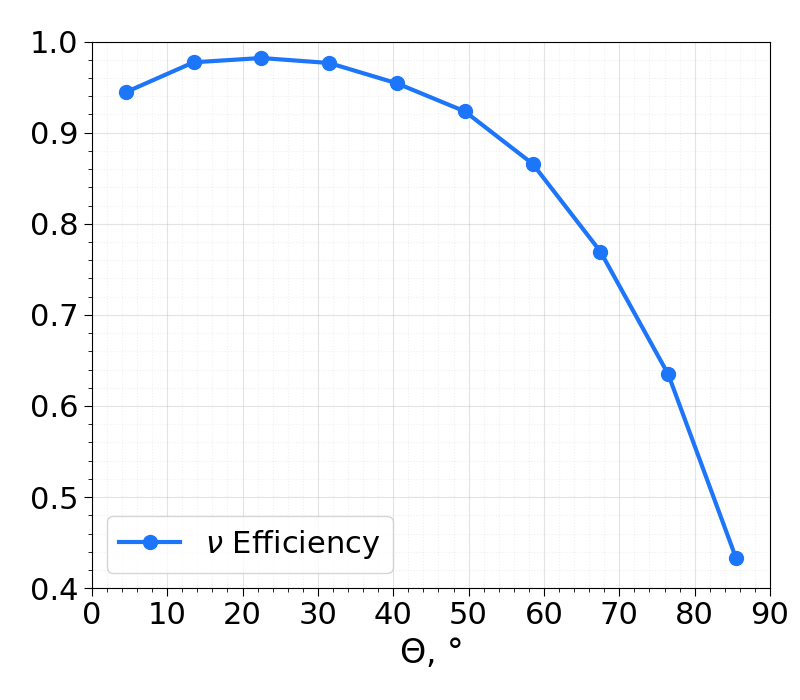}
    }\hfill
    \subcaptionbox{\label{fig:nu_classifier_eff_energy}}{%
        \includegraphics[width=0.32\textwidth]{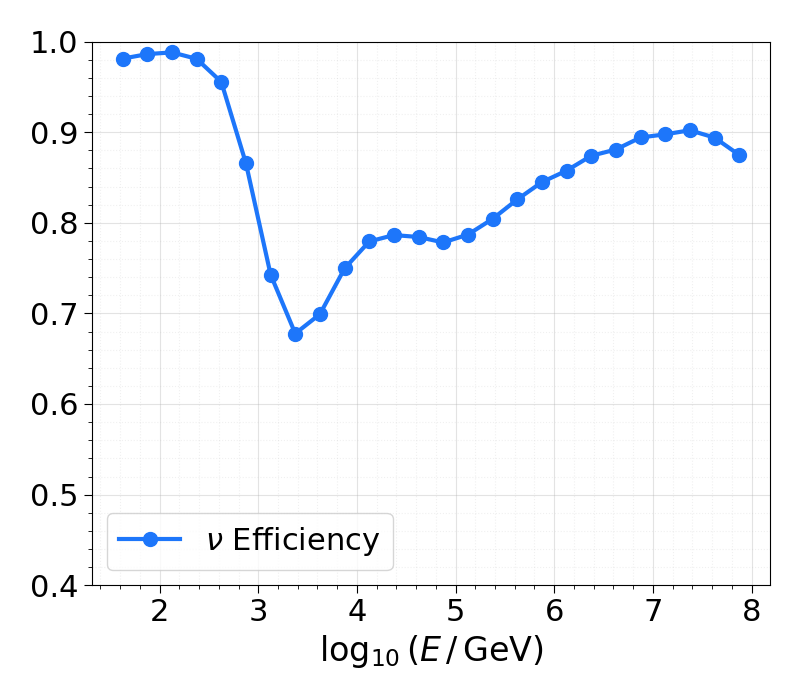}
    }
    \caption{
        \textit{Left:} EAS suppression factor as a function of $\nu_{\mu}$
        detection efficiency for the neutrino candidate extractor neural network, evaluated on a
        Monte Carlo test sample.
        The dotted horizontal line marks a suppression factor of $10^{6}$.
        \textit{Centre and right:} Neutrino selection efficiency as a function of the zenith arrival
        angle $\Theta$ and energy $\log_{10}(E/\mathrm{GeV})$, respectively.
        The efficiency is evaluated at for classification threshold corresponding to a suppression factor of $10^{6}$.
        The neutrino sample comprises equal-sized subsets of $\nu_\mu^{atm}$ and $\nu_\mu^{cosmo}$, with
        $N_{\nu} \approx 30{,}000$ events each, and $N_\mathrm{EAS} \approx 23{,}000{,}000$. In all panels, events are required to satisfy $n_{\text{hit}} \geq 8$ and $n_{\text{str}} \geq 3$.
    }
    \label{fig:nu_classifier_all}
\end{figure}

\textbf{Results.}
The model is evaluated on the MC test set after requiring at least 8 signal hits across at least 3 strings, as determined by the noise suppression model. Figure~\ref{fig:nu_classifier_SuppEff} shows the EAS suppression factor as a function of neutrino detection efficiency. At a classification threshold of $\xi = 0.9904$, the network achieves an EAS suppression factor of $10^{6}$ while retaining $83\%$ of neutrino events. This suppression factor matches the expected EAS-to-neutrino ratio in the Baikal-GVD data, meaning that at this working point the selected sample is expected to be dominated
by genuine neutrino-induced events rather than EAS background.

Figures \ref{fig:nu_classifier_eff_theta} and \ref{fig:nu_classifier_eff_energy} show the neutrino selection efficiency as a function of the zenith arrival angle $\Theta$ and primary particle energy, correspondingly. The zenith dependence reflects the fundamental difficulty of the classification problem: efficiency is high for clearly upgoing events ($\Theta \lesssim 60^\circ$), but falls steeply towards the horizon ($\Theta \to 90^\circ$), where the angular
separation between upgoing neutrinos and downgoing EAS is problematic. The energy dependence shows the same pattern as for the preliminary EAS suppression neural network, yet with lower efficiency values due to increased event classification threshold.

\begin{figure}[htbp]
    \centering
    \includegraphics[width=0.7\textwidth]{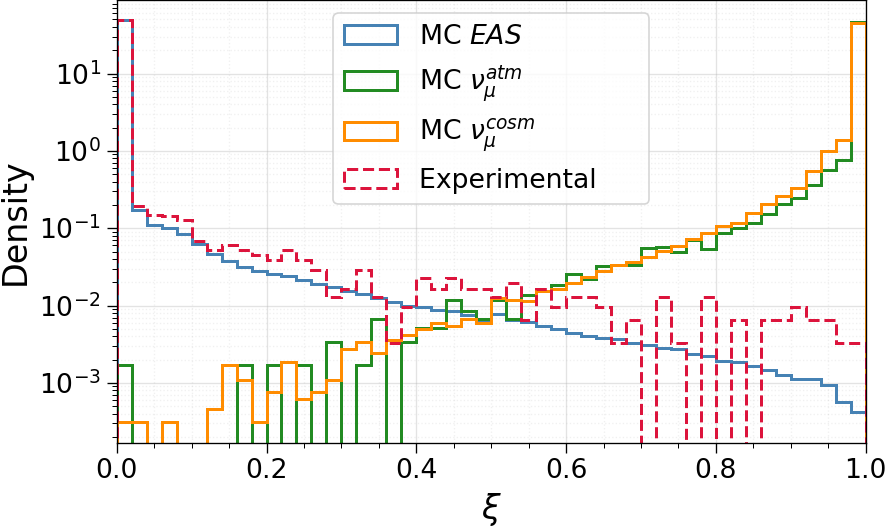}
        \caption{
    Distributions of the neutrino classifier output score $\xi$ for simulated EAS, 
    atmospheric $\nu_\mu$, cosmogenic $\nu_\mu$, and experimental events.
    All distributions are normalised to unit area.
    Events are required to have at least 8 signal hits across at least 3 signal strings.
    The displayed sample comprises $\approx 30{,}000$ atmospheric $\nu_\mu$ events, 
    $\approx 300{,}000$ cosmogenic $\nu_\mu$ events, $\approx 23{,}000{,}000$ EAS-induced events, 
    and $\approx 15{,}000$ experimental data events.
    }
    \label{fig:nu_classifier_score_distr}
\end{figure}

Figure~\ref{fig:nu_classifier_score_distr} shows the distribution of the classifier output score $\xi$ for MC EAS, atmospheric $\nu_\mu$, cosmogenic $\nu_\mu$, and experimental events. Neutrino-like events accumulate near $\xi = 1$, while EAS background events concentrate near $\xi = 0$, with a clear separation between the two classes. The MC EAS distribution agrees well with the experimental data over the broad score range $\xi \lesssim 0.8$, indicating that the Monte Carlo provides a reliable description of the dominant background.

In the high-score region $\xi > 0.8$, however, the experimental distribution begins to exceed the MC EAS prediction. Detailed investigation of these events reveals that a fraction of them arise from high-amplitude OM post-impulses~\cite{Haser:2013is} --- spurious late pulses following large primary signals --- which are not fully reproduced in the current MC simulation. Such events can be excluded by imposing event-quality cuts from the standard reconstruction procedure or by neural networks for further steps of data data analysis that are currently under development. We defer a complete treatment to future work. The presented network nonetheless extracts a high-purity neutrino candidate sample, and the MC--data agreement at intermediate scores confirms that the domain adaptation is effective in this stage of the pipeline.

%%%%%%%%%%%%%%%%%%%%
\section{Conclusion}
%%%%%%%%%%%%%%%%%%%%
\label{sec:conclusion}

We have presented a machine-learning-based data analysis pipeline for the Baikal-GVD neutrino telescope, comprising three transformer-based neural networks for EAS suppression, hit-level noise filtering, and neutrino candidate extraction. Applied sequentially, the pipeline achieves orders-of-magnitude acceleration of the data processing chain compared to the standard reconstruction-based approach, with increased signal efficiency and background rejection across all three tasks. The events identified as highly plausible neutrino candidates can be analyzed further to select pure set of neutrino events, significantly reducing time required for the analysis. Signal hits identified by the neural network at the second stage of data analysis can be used in all downstream tasks, providing increased efficiency of signal hits identification.

The domain shift between Monte Carlo simulations and experimental data is addressed throughout the pipeline via the domain-adversarial training technique \cite{Ganin:2016}. The effectiveness of this approach is confirmed quantitatively at each stage: for the EAS prefilter, domain adaptation reduces the Wasserstein distance between the simulated and experimental classifier output distributions from $W_1 = 0.0013$ to $W_1 = 0.0009$; for the noise suppression network, from $W_1 = 0.019$ to $W_1 = 0.008$, with the improvement visible in UMAP projections of the latent event representations. The consistent gains across all three stages demonstrate that domain-adversarial training is a robust and scalable strategy for bridging the simulation-to-data gap in water-Cherenkov neutrino detectors.

A notable methodological feature of the presented pipeline is that the neutrino candidate extractor is trained and operates on the output of the noise suppression network. This forms a stacked ML data analysis chain in which each stage conditions on the predictions of the preceding one. Such design relies critically on the reliability of the intermediate representations: if the noise filter were to produce outputs that differed systematically between MC and experimental data, the network for neutrino candidates identification would be exposed to a shifted input distributions, degrading its performance and reliability. The fact that it achieves consistent behavior on both MC and experimental data validates the entire chain, and demonstrates that a fully ML-based analysis workflow is achievable for Baikal-GVD.

The pipeline has been validated against experimental Baikal-GVD data and is under integration into the BARS data processing framework. Its inference speed makes it directly applicable to online event filtering and to the generation of multi-messenger alerts.

Looking ahead, we plan to develop a complete ML-based data selection and analysis chain that extends beyond the current pipeline. This includes reconstruction of energy and arrival direction of registered events, including the estimation of the corresponding uncertainty, and machine-learning-based event selection criteria. The ultimate goal is a full data analysis chain using neural networks, that can be used independently and in combination with other data analysis tools. Another perspective direction is to extend the presented pipeline to electron neutrino. Specifically, MC electron neutrino events are simulated in a separate event production chain and carry systematic differences with respect to muon neutrino samples. We plan to extend our pipeline to all neutrino types in future. 

%%%%%%%%%%%%%%%%%%%%
\section*{Acknowledgments}
%%%%%%%%%%%%%%%%%%%%
This work was supported by the Russian Science Foundation under grant no. 24-72-10056.

\appendix

%%%%%%%%%%%%%%%%%%%%
\section{Neural network architectures}
\label{app:nn_details}

In Table~\ref{table_nn_params} we present neural network architecture details used for training different tasks.

\begin{table}[htbp]
\centering
\begin{tabularx}{\textwidth}{|W{2.}|W{0.83}|W{0.83}|W{0.83}|W{0.83}|W{0.83}|W{0.83}|}
 \hline
Task & $N_\text{layers}$ & $N_\text{heads}$ & $d_\text{tr}$ & $d_\text{ff}$ & $N_\text{mlp}$ & $u_\text{mlp}$ \\
\hline
EAS suppression & 4 & 4 & 128 & 128 & 2 & [32,16] \\
\hline
Signal hits identification & 5 & 1 & 128 & 512 & 1 & 1 \\
\hline
$t_\text{res}$ estimation for signal hits & \multicolumn{4}{c|}{shared encoder} & 2 & 256 \\
\hline
Neutrino candidates search & 4 & 4 & 128 & 128 & 2 & [32, 16] \\
\hline
Domain adaptation head & -- & -- & -- & -- & 2 & [32,16]\\
\hline
\end{tabularx}

\caption{Neural network architecture details. $N_\text{layers}$ and $N_\text{heads}$ denote the number of transformer-encoder layers and attention heads, respectively; $d_\text{tr}$ and $d_\text{ff}$ are the embedding and feed-forward network dimensions; $N_\text{mlp}$ is the number of MLP layers in the prediction head; $u_\text{mlp}$ is the number of hidden units per MLP layer.}
\label{table_nn_params}
\end{table}

The signal hits identification and $t_\text{res}$ estimation neural network use ReLU and GELU activation functions, zero dropout, and the Adam optimizer with a target learning rate of $10^{-3}$ and batch size 128. An exponential warm-up schedule ramps the learning rate from zero to the target value over the first 100 training steps. After warm-up, a ReduceLROnPlateau scheduler halves the learning rate if the validation loss does not improve for 1000 steps, down to a minimum of $10^{-5}$.

Both the preliminary EAS suppression and the neutrino candidate extractor networks share the same architectural and optimisation set up. All layer use ReLU activation and do not use dropout. Both models are trained with AdamW (weight decay $10^{-2}$) optimizer and focal loss with $\gamma = 2$. A ReduceLROnPlateau scheduler halves the learning rate after 3 consecutive epochs without improvement, down to a minimum of $10^{-6}$, and training stops after 15 such epochs.
The preliminary EAS suppression model employs the learning rate of $10^{-3}$ for the feature extractor and classifier, and $2\times10^{-3}$ for the domain discriminator. For the neutrino candidate extractor network we set set reduced learning rate rates of $3\times10^{-4}$. Source (MC) train batches contain 512 events (EAS to neutrino events ratio 1:1) and target (experimental) batches have 256 events.

\bibliography{references}

@article{Gaisser:1995,
    author = {Gaisser, T. K. and Halzen, F. and Stanev, T.},
    title = {Particle astrophysics with high energy neutrinos},
    journal = {Phys. Rept.},
    volume = {258},
    pages = {173--236},
    year = {1995},
    doi = {10.1016/0370-1573(95)00003-Y}
}

@article{Halzen:2010,
    author = {Halzen, F. and Klein, S. R.},
    title = {IceCube: An Instrument for Neutrino Astronomy},
    journal = {Rev. Sci. Instrum.},
    volume = {81},
    pages = {081101},
    year = {2010},
    doi = {10.1063/1.3480478}
}

@article{Bartos:2020,
    author = {Bartos, I. and Kowalski, M.},
    title = {Multi-messenger astrophysics},
    journal = {Nature Rev. Phys.},
    volume = {2},
    pages = {446--458},
    year = {2020},
    doi = {10.1038/s42254-020-0222-4}
}

@article{Tamm:1937,
    author = {Tamm, I. and Frank, I. M.},
    title = {Coherent visible radiation of fast electrons passing through matter},
    journal = {C. R. Acad. Sci. URSS},
    volume = {14},
    pages = {109--114},
    year = {1937}
}

@article{Aartsen:2017jinst,
    author = {Aartsen, M. G. and others},
    collaboration = {IceCube},
    title = {The IceCube Neutrino Observatory: Instrumentation and Online Systems},
    journal = {JINST},
    volume = {12},
    pages = {P03012},
    year = {2017},
    doi = {10.1088/1748-0221/12/03/P03012}
}

@article{Adrian-Martinez:2016,
    author = {Adri{\'a}n-Mart{\'a}nez, S. and others},
    collaboration = {KM3NeT},
    title = {Letter of intent for KM3NeT 2.0},
    journal = {J. Phys. G},
    volume = {43},
    number = {8},
    pages = {084001},
    year = {2016},
    doi = {10.1088/0954-3899/43/8/084001}
}

@article{Avrorin:2019,
    author = {Avrorin, A. D. and others},
    collaboration = {Baikal-GVD},
    title = {Baikal-{GVD}: the Next Generation Neutrino Telescope in {L}ake {B}aikal},
    journal = {J. Phys. Conf. Ser.},
    volume = {1263},
    pages = {012005},
    year = {2019},
    doi = {10.1088/1742-6596/1263/1/012005}
}

@article{Safronov:2020,
    author = {Safronov, Grigory},
    collaboration = {Baikal-GVD},
    title = {{Baikal-GVD}: status and first results},
    journal = {PoS},
    volume = {ICHEP2020},
    pages = {606},
    year = {2020},
    doi = {10.22323/1.390.0606},
    note = {arXiv:2012.03373}
}

@article{Safronov:2021EPJC,
    author = {Safronov, G. and others},
    collaboration = {Baikal-GVD},
    title = {Measuring muon tracks in {Baikal-GVD} using a fast reconstruction algorithm},
    journal = {Eur. Phys. J. C},
    volume = {81},
    pages = {1025},
    year = {2021},
    doi = {10.1140/epjc/s10052-021-09825-y}
}

@article{Allakhverdyan:2021,
    author = {Allakhverdyan, V. A. and others},
    collaboration = {Baikal-GVD},
    title = {Multi-messenger and real-time astrophysics with the {Baikal-GVD} telescope},
    journal = {PoS},
    volume = {ICRC2021},
    pages = {0946},
    year = {2021},
    doi = {10.22323/1.395.0946}
}

@article{Allakhverdyan:2023diffuse,
    author = {Allakhverdyan, V. A. and others},
    collaboration = {Baikal-GVD},
    title = {Diffuse neutrino flux measurements with the {Baikal-GVD} neutrino telescope},
    journal = {Phys. Rev. D},
    volume = {107},
    pages = {042005},
    year = {2023},
    doi = {10.1103/PhysRevD.107.042005}
}

@article{Avrorin:2016OM,
    author = {Avrorin, A. D. and others},
    collaboration = {Baikal-GVD},
    title = {The optical module of {Baikal-GVD}},
    journal = {Phys. Part. Nucl. Lett.},
    volume = {13},
    pages = {737--746},
    year = {2016},
    doi = {10.1134/S1547477116060029}
}

@article{Avrorin:2002groupvelocity,
    author = {Avrorin, A. D. and others},
    collaboration = {Baikal},
    title = {Measurements of group velocity of light in the {L}ake {B}aikal water},
    journal = {Nucl. Instrum. Meth. A},
    volume = {480},
    pages = {737--742},
    year = {2002},
    doi = {10.1016/S0168-9002(01)01241-4}
}

@article{Avrorin:2019noise,
    author = {Avrorin, A. D. and others},
    collaboration = {Baikal-GVD},
    title = {The optical noise monitoring systems of {L}ake {B}aikal environment for the {Baikal-GVD} telescope},
    journal = {PoS},
    volume = {ICRC2019},
    pages = {875},
    year = {2019},
    doi = {10.22323/1.358.0875},
    note = {arXiv:1908.06509}
}

@article{Rjabov:2021,
    author = {Rjabov, E. V. and Tarashansky, B.},
    collaboration = {Baikal-GVD},
    title = {Monitoring of optical properties of deep lake water},
    journal = {PoS},
    volume = {ICRC2021},
    pages = {1034},
    year = {2021},
    doi = {10.22323/1.395.1034}
}

@article{Dvornicky:2021,
    author = {Dvornick{\'y}, R. and others},
    collaboration = {Baikal-GVD},
    title = {The {Baikal-GVD} neutrino telescope as an instrument for studying {B}aikal water luminescence},
    journal = {PoS},
    volume = {ICRC2021},
    pages = {1113},
    year = {2021},
    doi = {10.22323/1.395.1113}
}

@article{Kharuk:2024,
    author = {Kharuk, I. and others},
    title = {Application of Machine Learning Methods in {Baikal-GVD}: Background Noise Rejection and Selection of Neutrino-Induced Events},
    journal = {Moscow Univ. Phys. Bull.},
    volume = {79},
    pages = {97--103},
    year = {2024},
    doi = {10.3103/S0027134924700188}
}

@article{Dik:2025,
    author = {Dik, V. and Suvorova, O.},
    collaboration = {Baikal-GVD},
    title = {Online Analysis and Multimessenger Alerts Follow-Up at the {Baikal-GVD} Telescope},
    journal = {PoS},
    volume = {ICRC2025},
    pages = {921},
    year = {2025},
    doi = {10.22323/1.501.0921}
}

@article{Aartsen:2013science,
    author = {Aartsen, M. G. and others},
    collaboration = {IceCube},
    title = {Evidence for High-Energy Extraterrestrial Neutrinos at the IceCube Detector},
    journal = {Science},
    volume = {342},
    number = {6161},
    pages = {1242856},
    year = {2013},
    doi = {10.1126/science.1242856}
}

@article{Aartsen:2018blazar,
    author = {Aartsen, M. G. and others},
    collaboration = {IceCube},
    title = {Multimessenger observations of a flaring blazar coincident with high-energy neutrino {I}ce{C}ube-170922{A}},
    journal = {Science},
    volume = {361},
    number = {6398},
    pages = {eaat1378},
    year = {2018},
    doi = {10.1126/science.aat1378}
}

@incollection{Heck:1998,
    author = {Heck, D. and Knapp, J. and Capdevielle, J. N. and Schatz, G. and Thouw, T.},
    title = {CORSIKA: A Monte Carlo code to simulate extensive air showers},
    booktitle = {Report FZKA 6019},
    publisher = {Forschungszentrum Karlsruhe},
    year = {1998}
}

@article{Ostapchenko:2006,
    author = {Ostapchenko, S.},
    title = {QGSJET-II: towards reliable description of very high energy hadronic interactions},
    journal = {Nucl. Phys. B Proc. Suppl.},
    volume = {151},
    pages = {143--146},
    year = {2006},
    doi = {10.1016/j.nuclphysbps.2005.07.026}
}

@inproceedings{Vaswani:2017,
    author = {Vaswani, Ashish and Shazeer, Noam and Parmar, Niki and Uszkoreit, Jakob and Jones, Llion and Gomez, Aidan N. and Kaiser, {\L}ukasz and Polosukhin, Illia},
    title = {Attention Is All You Need},
    booktitle = {Advances in Neural Information Processing Systems},
    volume = {30},
    year = {2017},
    url = {https://papers.nips.cc/paper/7181-attention-is-all-you-need}
}

@article{Ganin:2016,
    author = {Ganin, Yaroslav and Ustinova, Evgeniya and Ajakan, Hana and Germain, Pascal and Larochelle, Hugo and Laviolette, Fran{\c{c}}ois and Marchand, Mario and Lempitsky, Victor},
    title = {Domain-Adversarial Training of Neural Networks},
    journal = {J. Mach. Learn. Res.},
    volume = {17},
    number = {59},
    pages = {1--35},
    year = {2016},
    url = {http://jmlr.org/papers/v17/15-239.html}
}

@inproceedings{Alon:2021,
    author = {Alon, Uri and Yahav, Eran},
    title = {On the Bottleneck of Graph Neural Networks and its Practical Implications},
    booktitle = {International Conference on Learning Representations (ICLR)},
    year = {2021},
    url = {https://arxiv.org/abs/2006.05205}
}

@inproceedings{Devlin:2019,
    author = {Devlin, Jacob and Chang, Ming-Wei and Lee, Kenton and Toutanova, Kristina},
    title = {{BERT}: Pre-training of Deep Bidirectional Transformers for Language Understanding},
    booktitle = {Proceedings of the 2019 Conference of the North American Chapter of the Association for Computational Linguistics: Human Language Technologies},
    pages = {4171--4186},
    year = {2019},
    url = {https://aclanthology.org/N19-1423}
}

@article{allakhverdyan2021efficient,
  title={An efficient hit finding algorithm for Baikal-GVD muon reconstruction},
  author={Allakhverdyan, VA and Avrorin, AD and Avrorin, AV and Aynutdinov, VM and Bannasch, R and Barda{\v{c}}ov{\'a}, Z and Belolaptikov, IA and Borina, IV and Brudanin, VB and Budnev, NM and others},
  journal={arXiv preprint arXiv:2108.00208},
  year={2021}
}

@misc{McInnes2018,
      title={UMAP: Uniform Manifold Approximation and Projection for Dimension Reduction}, 
      author={Leland McInnes and John Healy and James Melville},
      year={2020},
      eprint={1802.03426},
      archivePrefix={arXiv},
      primaryClass={stat.ML},
      url={https://arxiv.org/abs/1802.03426}, 
}

@inproceedings{lin2017focal,
  title={Focal loss for dense object detection},
  author={Lin, Tsung-Yi and Goyal, Priya and Girshick, Ross and He, Kaiming and Doll{\'a}r, Piotr},
  booktitle={Proceedings of the IEEE international conference on computer vision},
  pages={2980--2988},
  year={2017}
}

@article{Haser:2013is,
    author = {Haser, J. and Kaether, F. and Langbrandtner, C. and Lindner, M. and Lucht, S. and Roth, S. and Schumann, M. and Stahl, A. and St{\"u}ken, A. and Wiebusch, C.},
    title = "{Afterpulse Measurements of R7081 Photomultipliers for the Double Chooz Experiment}",
    eprint = "1301.2508",
    archivePrefix = "arXiv",
    primaryClass = "physics.ins-det",
    doi = "10.1088/1748-0221/8/04/P04029",
    journal = "JINST",
    volume = "8",
    pages = "P04029",
    year = "2013"
}

%%%%%%%%%%%%%%%%%%%%
\end{document}